\documentclass[10pt,twocolumn,letterpaper]{article}

\usepackage{iccv}              

\definecolor{iccvblue}{rgb}{0.21,0.49,0.74}
\usepackage[pagebackref,breaklinks,colorlinks,allcolors=iccvblue]{hyperref}

\DeclareMathOperator*{\argmax}{arg\,max}
\DeclareMathOperator*{\argmin}{arg\,min}

\usepackage{rotating}
\usepackage{multirow}
\usepackage{float}
\usepackage{amsmath}
\usepackage{orcidlink}
\usepackage[accsupp]{axessibility}  
\usepackage[bottom]{footmisc}

\def\paperID{9589} 
\def\confName{ICCV}
\def\confYear{2025}

\title{When and Where do Data Poisons Attack Textual Inversion?}

\author{Jeremy Styborski$^{1*}$ \ \ \ 
Mingzhi Lyu$^{2*}$ \ \ \ 
Jiayou Lu$^1$ \ \ \ 
Nupur Kapur$^1$ \ \ \
Adams Wai-Kin Kong$^1$ \\ 
$^1$College of Computing and Data Science, Nanyang Technological University, Singapore \\
$^2$Rapid-Rich Object Search (ROSE) Lab, Nanyang Technological University, Singapore \\
{\tt\small{\{styb0001,lyum0002,jiayou001,nupur003,AdamsKong\}@ntu.edu.sg}}}

\begin{document}
\maketitle
\def\thefootnote{\tt\normalsize*}\footnotetext{Equal contribution, may cite either first}
\renewcommand*{\thefootnote}{\arabic{footnote}}
\begin{abstract}


Poisoning attacks pose significant challenges to the robustness of diffusion models (DMs). In this paper, we systematically analyze when and where poisoning attacks textual inversion (TI), a widely used personalization technique for DMs. We first introduce Semantic Sensitivity Maps, a novel method for visualizing the influence of poisoning on text embeddings. Second, we identify and experimentally verify that DMs exhibit non-uniform learning behavior across timesteps, focusing on lower-noise samples. Poisoning attacks inherit this bias and inject adversarial signals predominantly at lower timesteps. Lastly, we observe that adversarial signals distract learning away from relevant concept regions within training data, corrupting the TI process. Based on these insights, we propose Safe-Zone Training (SZT), a novel defense mechanism comprised of 3 key components: (1) JPEG compression to weaken high-frequency poison signals, (2) restriction to high timesteps during TI training to avoid adversarial signals at lower timesteps, and (3) loss masking to constrain learning to relevant regions. Extensive experiments across multiple poisoning methods demonstrate that SZT greatly enhances the robustness of TI against all poisoning attacks, improving generative quality beyond prior published defenses. 

\noindent\textbf{Code:} \url{www.github.com/JStyborski/Diff_Lab}

\noindent\textbf{Data:} \url{www.github.com/JStyborski/NC10}

\end{abstract}    
\section{Introduction}
\label{sec:intro}

The image quality and prompt fidelity offered by diffusion-based image generation models such as DALL-E 2 \cite{2022_Ramesh_DallE2}, Stable Diffusion \cite{2022_Rombach_LDM}, and Imagen \cite{2022_Saharia_Imagen} have popularized (or vilified) the use of AI-generated images in the art, marketing, and media industries. The subsequent proliferation of personalization and editing methods \cite{2023_Ruiz_DreamBooth, 2022_Gal_TextInv, 2021_Hu_LoRA, 2023_Kumari_CustomDiffusion, 2022_Hertz_Prompt2Prompt, 2023_Zhang_ControlNet} and tools \cite{2022_AUTOMATIC1111_SDWebUI, 2022_Von-Platen_Diffusers, 2025_??_CivitAI} allows anyone, even non-experts and non-artists, to quickly retrain existing models to generate their desired images, including images of novel concepts not found in the model training data.

Unfortunately, the surge in generative AI corresponds to misfortune for artists and copyright owners, who may see their works easily reproduced by retrained generative models. Consequently, the subfield of data poisoning for generative AI has focused on how to inject images with signals that nullify their use in training datasets for generative models. Individuals looking to protect their data are encouraged to ``poison" their images with imperceptible adversarial perturbations such that generative models will fail to learn to recreate the novel concepts contained within.

\begin{figure}[t]
  \centering
  \includegraphics[width=\linewidth,height=0.65\linewidth]{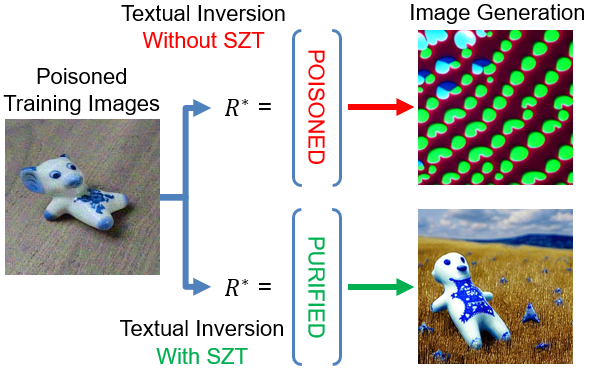}
  \caption{Applying SZT to TI mitigates poison signals and recovers shape, color, and texture of desired concepts. NovelConcepts10 training data poisoned by ADM+. Generation prompt: ``A $R*$ with a wheat field in the background"}
  \label{fig:SZT_Example}
  \vspace{-0.5cm}
\end{figure}

The concept of data poisoning stems from adversarial examples (AEs) \cite{2014_Szegedy_PreFGSM, 2015_Goodfellow_FGSM, 2017_Kurakin_PGD, 2019_Madry_PGD}, where the inputs are slightly perturbed in order to maximize loss. Intuitively, these AEs learn to incorporate signals that maximally mislead AI models into making incorrect predictions. The AE method is likewise effective against generative models; multiple state-of-the-art poisoning methods for generative models apply the AE method to the diffusion model (DM) objective to inject poison signals into images \cite{2023_Salman_Photoguard, 2023_Liang_AdvDM, 2024_Xue_SDS, 2025_Choi_DiffusionGuard}. Although the textual inversion (TI) technique is widely used to adapt DMs to novel concepts, there are no systematic studies that identify when and where AEs attack the TI process.

\subsection{Research Gap and Contributions}

State-of-the-art poisoning methods for DMs are largely cannibalizing previous adversarial methods (e.g., Glaze \cite{2023_Shan_Glaze} from Adv-VAE \cite{2016_Tabacof_AdvVAE} or AdvDM \cite{2023_Liang_AdvDM} from PGD \cite{2017_Kurakin_PGD, 2019_Madry_PGD}) or creating new loss functions that nonetheless follow the same AE algorithm (e.g., SDS \cite{2024_Xue_SDS} or DiffusionGuard \cite{2025_Choi_DiffusionGuard}). Although these poisoning methods are useful research milestones that explore novel applications for AEs, they neglect to study how poison signals are learned and which DM vulnerabilities are being exploited. Furthermore, many poisoning works neglect to understand the intricacies of the downstream tasks, such as TI, for which the AEs are designed. 

In this paper, we examine the learning behavior of TI \cite{2022_Gal_TextInv}, a lightweight personalization technique for adapting DMs to novel concepts, when trained on poisoned datasets. To enable our analysis, we devise a novel visualization method, semantic sensitivity maps (SSM), to display loss sensitivity to individual text tokens throughout training. Using SSMs, we observe several heretofore unseen trends in poison learning: (1) DMs concentrate poison learning at lower-middle timesteps (approximately $t \in [0,600]$) and (2) adversarial signals tend to ``distract" learning away from the novel concept of interest and towards irrelevant regions.

We scrutinize the noising process and the noise-prediction training objective that are common to DMs and the TI process to demonstrate that their combined effect inherently favors low-noise images. Crucially, this learning bias towards lower-middle timesteps is inherent in most AE algorithms for DMs, which implies that the AEs themselves are mostly learned at lower-middle timesteps. This exposes a weakness where many poisoning techniques are less effective at higher timesteps. In experiments, we demonstrate that restricting TI training to higher timesteps is a simple yet effective defense against multiple poisons.

We find that TI prioritizes learning in regions that contain concept objects rather than background objects. This behavior stems from the tendency to minimize noise-prediction loss by focusing on common features across training images. Poisoning disrupts this process by introducing dominant adversarial signals and suppressing the learnability of clean concept information, misleading the TI training process. To counteract this, we propose a loss-masking strategy during optimization, explicitly training the model to focus on the concept regions within poisoned images. This ensures that TI effectively captures the intended concept and mitigates the influence of adversarial signals.

We also analyze JPEG compression, a common defense technique in poisoning literature. We find that JPEG compression helps purify AEs by reshaping the distributions of adversarial signal magnitudes and frequencies to benign distributions, similar to those of clean images. Although JPEG compression alone cannot enable TI to learn from poisoned images, using it as a first step to weaken high-frequency adversarial signals significantly improves the performance of other defense methods.

Finally, we combine our analyses to present safe-zone training (SZT), a simple and lightweight defense method for training TI. SZT consists of three main components: (1) JPEG compression to weaken high-frequency adversarial signals, (2) training at high timesteps to avoid adversarial signals at lower-middle timesteps, and (3) loss masking to prioritize learning on the novel concept. We show that SZT outperforms multiple established defenses across six different poisons in accurately generating novel concepts and reaches the performance of TI on clean data. 

Our contributions are summarized below:
\begin{itemize}
  \item We devise SSMs and use them to visualize loss attribution during the TI training process.
  \item We analyze the noising process and noise-prediction objective to understand the mechanism by which DMs, AEs, and TI focus on lower-middle timesteps.
  \item We develop a loss-masking strategy to encourage DMs to learn from the concept regions in the training data.
  \item We combine our findings and establish SZT to methodically improve the poison robustness of TI.
\end{itemize}

\section{Related Works}
\label{sec:litreview}

We assume that the reader is familiar with standard DMs \cite{2020_Ho_DDPM, 2020_Song_Score, 2021_Song_SDE, 2023_Kingma_VDM} and generation techniques \cite{2022_Song_DDIM, 2022_Meng_SDEdit, 2021_Dhariwal_DDIMInv, 2022_Ho_CFG}. We mainly focus on latent diffusion models (LDMs) \cite{2022_Rombach_LDM}.

Although DMs can easily generate most concepts from their training dataset with high quality, they predictably suffer in generation quality for novel (unseen) concepts. Editing techniques \cite{2022_Meng_SDEdit, 2022_Hertz_Prompt2Prompt, 2022_Mokady_NullTextInv, 2024_Miyake_NegPromptInv, 2024_Ju_PnPInversion, 2023_Zhang_ControlNet, 2023_Brooks_InstructPix2Pix} apply modifications to an input image while enforcing structural or composition constraints, permitting generation with novel concept images. However, editing techniques are still bound to the prior from the DM's training data. In order to ``inject" new concepts and styles into the DM, personalization techniques \cite{2021_Hu_LoRA, 2022_Gal_TextInv, 2023_Han_SVDiff, 2024_Zhang_CompInversion, 2023_Wei_ELITE, 2024_Motamed_Lego, 2023_Ruiz_DreamBooth, 2023_Kumari_CustomDiffusion, 2023_Voynov_XTI, 2023_Gu_MixofShow} quickly finetune some component of the DM on a small dataset of novel concept images. Many personalization techniques learn new text embeddings \cite{2022_Gal_TextInv, 2023_Gu_MixofShow, 2023_Kumari_CustomDiffusion, 2023_Voynov_XTI, 2023_Wei_ELITE} or co-opt existing tokens \cite{2023_Ruiz_DreamBooth, 2023_Han_SVDiff} to capture concept semantics. Without retraining the entire DM from scratch, personalization techniques are the best way to adapt DMs to new concepts. In particular, TI \cite{2022_Gal_TextInv} is a lightweight personalization method that requires training only one new text embedding and is often included within other methods \cite{2023_Kumari_CustomDiffusion, 2023_Voynov_XTI, 2023_Gu_MixofShow}.

AEs \cite{2014_Szegedy_PreFGSM, 2015_Goodfellow_FGSM, 2017_Carlini_CarliniWagner, 2017_Kurakin_PGD, 2019_Madry_PGD} expose the brittleness of neural net predictions by learning minor perturbations to maximize loss. Although originally designed for classification models, AEs have also been applied to object detection and segmentation \cite{2017_Xie_AEObjDet}, reinforcement learning \cite{2017_Huang_AE4RL}, watermarking \cite{2023_Lyu_AdvWatermark}, self-supervised learning \cite{2020_Kim_RoCL, 2020_Jiang_ACL, 2022_Carlini_CLIPPoison}, VAEs \cite{2016_Tabacof_AdvVAE, 2017_Kos_AdvGen}, and diffusion models \cite{2023_Shan_Glaze, 2023_Le_AntiDB, 2023_Liang_AdvDM, 2024_Shan_Nightshade, 2024_Xue_SDS, 2023_Salman_Photoguard, 2025_Choi_DiffusionGuard}. Due to their ability to defeat deep learning models, AEs have become a landmark technique for data protection. The concept of AEs extends to adversarial poisoning \cite{2021_Fowl_TAP, 2021_Huang_UE}, where entire datasets are poisoned to nullify attempts to scrape or steal data. Typical defenses against AEs include simple detection \cite{2022_Aldahdooh_AEDetector}, augmentation \cite{DeVries_2017_Cutout, Zhang_2018_Mixup, Yun_2019_Cutmix, Cubuk_2019_RandAugment, 2019_Cohen_GaussNoise, 2023_Liu_ISS}, which modifies input images to increase sample diversity and dilute poison signals, multitask architectures \cite{2023_Xue_Combined_Theory, 2024_Styborski_VESPR}, which use multiple networks to mitigate shortcut learning \cite{2018_Geirhos_Texture_Shortcuts, 2021_Lyu_Simplicity_Bias}, and adversarial training, which simply adds AEs to the training set.

AEs for DMs fall into two main categories. Encoder attacks \cite{2023_Shan_Glaze, 2024_Shan_Nightshade, 2023_Salman_Photoguard} utilize the encoder of an LDM to learn a small perturbation to an input image that minimizes the distance between the perturbed image encoding and the encoding of some unrelated target image. Diffuser attacks \cite{2023_Le_AntiDB, 2023_Liang_AdvDM, 2024_Xue_SDS, 2023_Salman_Photoguard, 2025_Choi_DiffusionGuard} aim to learn a small perturbation to an input image that maximizes the DM loss (often noise-prediction). Other attack methods combine encoder attacks and diffuser attacks \cite{2023_Liang_Mist, 2023_Salman_Photoguard}. Defenses against AEs for DMs include augmentation \cite{2025_Shidoto_AdverseCleaner} and regeneration \cite{2024_Dolatabadi_AVATAR, 2024_Xue_PDMPure, 2024_Zhao_Regeneration, 2024_Liu_RemovingWatermarks}, where poisoned images are randomly noised and then denoised with a DM to remove adversarial signals. Regeneration methods are the most actively researched, but they are time-consuming and complex as they require preprocessing entire datasets with a DM in order to ``purify" them.

\section{Preliminaries}
\label{sec:preliminaries}

\subsection{Latent Diffusion Models}

We mainly focus on LDMs \cite{2022_Rombach_LDM} as they are quick to train, they generate high-quality outputs, and tools for using/modifying them are widely available. In LDMs, an input image $x$ is first encoded to a latent of lower dimension via a variational encoder, $z_{0}=\mathcal{E}(x)$\footnote{We abuse notation and let $\mathcal{E}(x)$ encompass encoding and sampling.}. The latent is then iteratively noised to timestep $t$ according to noise variance schedule $\beta_{t}$. An equivalent one-step noising process is
\begin{equation}
    \label{eqn:Noising}
    z_{t} = \sqrt{\overline{\alpha}_{t}}z_{0} + \sqrt{1-\overline{\alpha}_{t}}\epsilon,
\end{equation}
where $\overline{\alpha}_{t}=\Sigma_{s=1}^t (1-\beta_{s})$ and $\epsilon \sim \mathcal{N}(\textbf{0}, I)$. $\overline{\alpha}_{t}$ monotonically decreases from $1$ to $0$ as $t$ increases from $0$ to $T$, following a variance-preserving schedule \cite{2021_Song_SDE}. A generative DM learns the denoising process, where the goal is to maximize the likelihood of generating the unnoised input ($z_{0}$). DDPM \cite{2020_Ho_DDPM} showed that this objective can be practically realized with a simple noise-prediction task. The model predicts the Gaussian noise $\epsilon$ added to input $z_{0}$ when given noisy sample $z_{t}$, and the learning objective is given as
\begin{equation}
    \label{eqn:Noise_Pred}
    \argmin_\theta \ E_{x,t \sim \mathcal{U}(0,T),\epsilon}\left[||\epsilon_{\theta}(z_{t},t)-\epsilon||_{2}^2\right],
\end{equation}
where $x$ is drawn from the training dataset, $\mathcal{U}(0,T)$ is the uniform distribution between timesteps $0$ and $T$, and $\epsilon_{\theta}$ is a model that predicts a noise vector $\epsilon$ given a noised latent $z_{t}$. For text-conditional models, $\epsilon_{\theta}(z_{t},t,\tau_{\theta}(y))$ also accepts features vector $\tau_{\theta}(y)$, where $\tau_{\theta}$ is a language model and input prompt $y$ is provided by the user or training data.

\subsection{Textual Inversion}

In this paper, we consider how AEs affect the downstream personalization task of TI \cite{2022_Gal_TextInv}, which seeks to learn a new text token that captures the semantics of a novel concept. TI trains a new text embedding corresponding to a new token $R^*$; the new embedding is the only finetuned parameter. TI trains via a noise-prediction task as
\begin{equation}
    \label{eqn:Textual_Inversion}
    \argmin_{\theta_{R^*}} E_{x_{f},t \sim \mathcal{U}(0,T),y^*,\epsilon}\left[||\epsilon_{\theta}(z_{t},t,\tau_{\theta}(y^*))-\epsilon||_{2}^2\right],
\end{equation}
where sample $x_{f}$ is drawn from the finetuning dataset and $y^*$ is a prompt containing token $R^*$. 

\subsection{Adversarial Examples}

The primary method to create AEs against LDMs is by attacking the diffuser with an adversarial perturbation $\delta$ injected into the input data. The adversarial objective is to find a $\delta$ that maximizes the noise-prediction loss, as
\begin{equation}
    \label{eqn:Diffusion_Attack}
    \begin{split}
    \argmax_\delta & \ E_{x_{p},t \sim \mathcal{U}(0,T),\epsilon}\left[||\epsilon_{\theta}(z'_{t},t)-\epsilon||_{2}^2\right] \\
    & \text{s.t.} \quad ||\delta||_b  \leq \kappa,
    \end{split}
\end{equation}
where sample $x_{p}$ is drawn from the dataset to poison and $z'_{t}=\sqrt{\overline{\alpha}_{t}}\mathcal{E}(x_{p}+\delta) + \sqrt{1-\overline{\alpha}_{t}}\epsilon$ is the adversarial latent following Eq. \ref{eqn:Noising}. The perturbation $\delta$ is commonly constrained in the $l^\infty$ norm to a small value $\kappa$ to limit its visibility.

Encoder attacks against LDMs seek to perturb an image such that the encoding of the poisoned image $x_{p}+\delta$ is similar to that of some unrelated target image $x_{g}$. Encoder attacks optimize the objective function
\begin{equation}
    \label{eqn:Encoder_Attack}
    \argmin_\delta \ E_{x_{p}} \left[ ||\mathcal{E}(x_{p}+\delta)-\mathcal{E}(x_{g})||_{2}^2 \right] \ \text{s.t.} \ ||\delta||_b \leq \kappa.
\end{equation}

\section{Analysis}
\label{sec:Analysis}

\subsection{Visualizing the Effect of Adversarial Examples}
\label{subsec:Semantic_Sensitivity_Map}

To observe the influence of AEs against TI, it is critical to visualize the impact of each text embedding on a generated image. Let $\widehat{e}=\{e_0, ... e_n, ..., e_{L-1}\}$, where $e_0$ represents the start token embedding vector and $e_n$ represents the embedding vector of the $n$th token. For a text input consisting of $l$ tokens, $\widehat{e}$ is padded to length $L$ with $\{e_{l+1}, ..., e_{L-1}\}$ with end token embeddings. Each $e_n$ is obtained via lookup in a token-embedding dictionary. As an example, given the text prompt, ``a puppy wearing a hat", embeddings $\{e_1,...,e_5\}$ correspond to ``a", ``puppy", ``wearing", ``a", and ``hat" respectively. 

\begin{figure}[t]
  \centering
  \includegraphics[width=0.9\linewidth]{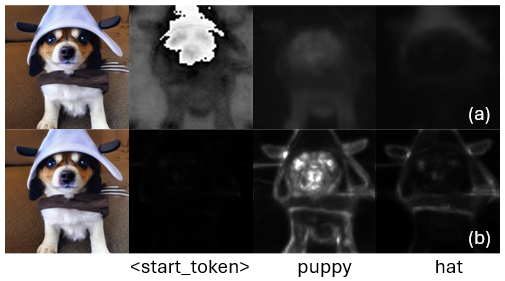}
  \caption{Cross attention maps (a) and SSMs (b) for tokens ``\textless start \textunderscore token\textgreater", ``puppy", and ``hat" with text prompt ``a puppy wearing a hat" at noise timestep $t=500$.}
  \label{fig:cross_attentionVSsemantic}
  \vspace{-0.5cm}
\end{figure}

A popular text-attribution method for LDMs visualizes the cross-attention maps \cite{2024_Liu_Attention} in $\epsilon_\theta$. However, cross-attention maps are incapable of capturing the effect of each embedding $e_n$. This is because the self-attention layers in the text encoder $\tau_{\theta}$ entangle signals from the embedding vectors of $\widehat{e}$ before they reach the cross-attention layers of $\epsilon_\theta$. Fig. \ref{fig:cross_attentionVSsemantic}(a) visualizes text attribution using cross-attention maps. The cross-attention map corresponding to the start token $e_{0}$ is bright, highlighting many regions, but the maps corresponding to words ``puppy" and ``hat" are dim and blurry. This drawback has been noted in other works \cite{2023_Lu_TF-Icon}.

To accurately identify the spatial regions over which a specific token has influence, we developed a new visualization method called semantic sensitivity maps (SSMs). We define SSMs by
\begin{align}
    \label{eqn:sensitivity map}
    SSM (x, t, \widehat{e}, n) & = E_{\widehat{e}_{\Delta n}} \left[ ||\epsilon_{\theta}(z_{t},t,c) - \epsilon_{\theta}(z_{t},t,c_{\Delta n})||_{ch}^2 \right ] \notag \\
    \text{s.t.} \ c & = \tau_{\theta}(\widehat{e}), \ c_{\Delta n} = \tau_{\theta}(\widehat{e}_{\Delta n}),
\end{align}
where $x$ is an input image and $c$ is the condition tensor. $c_{\Delta n}$ is analogous to $c$, except that it is derived from $\widehat{e}_{\Delta n}$, a modified version of $\widehat{e}$ that replaces $e_n$ with a randomly sampled vector from the embedding dictionary. For notational simplicity, we denote that $\tau_{\theta}(\widehat{e})$ accepts token embeddings as input instead of the prompt. $||\cdot||_{ch}^2$ indicates sum of squares across channels such that the output is a one-channel image that matches the input height and width. Fig. \ref{fig:cross_attentionVSsemantic}(b) displays the SSMs corresponding to the embeddings for ``puppy" and ``hat". Compared to cross-attention attribution maps in Fig. \ref{fig:cross_attentionVSsemantic}(a), SSMs more accurately reflect the regions corresponding to each input embedding. Crucially, SSMs avoid attributing semantic tokens to the start token.

\begin{figure}[t]
  \centering
  \includegraphics[width=\linewidth]{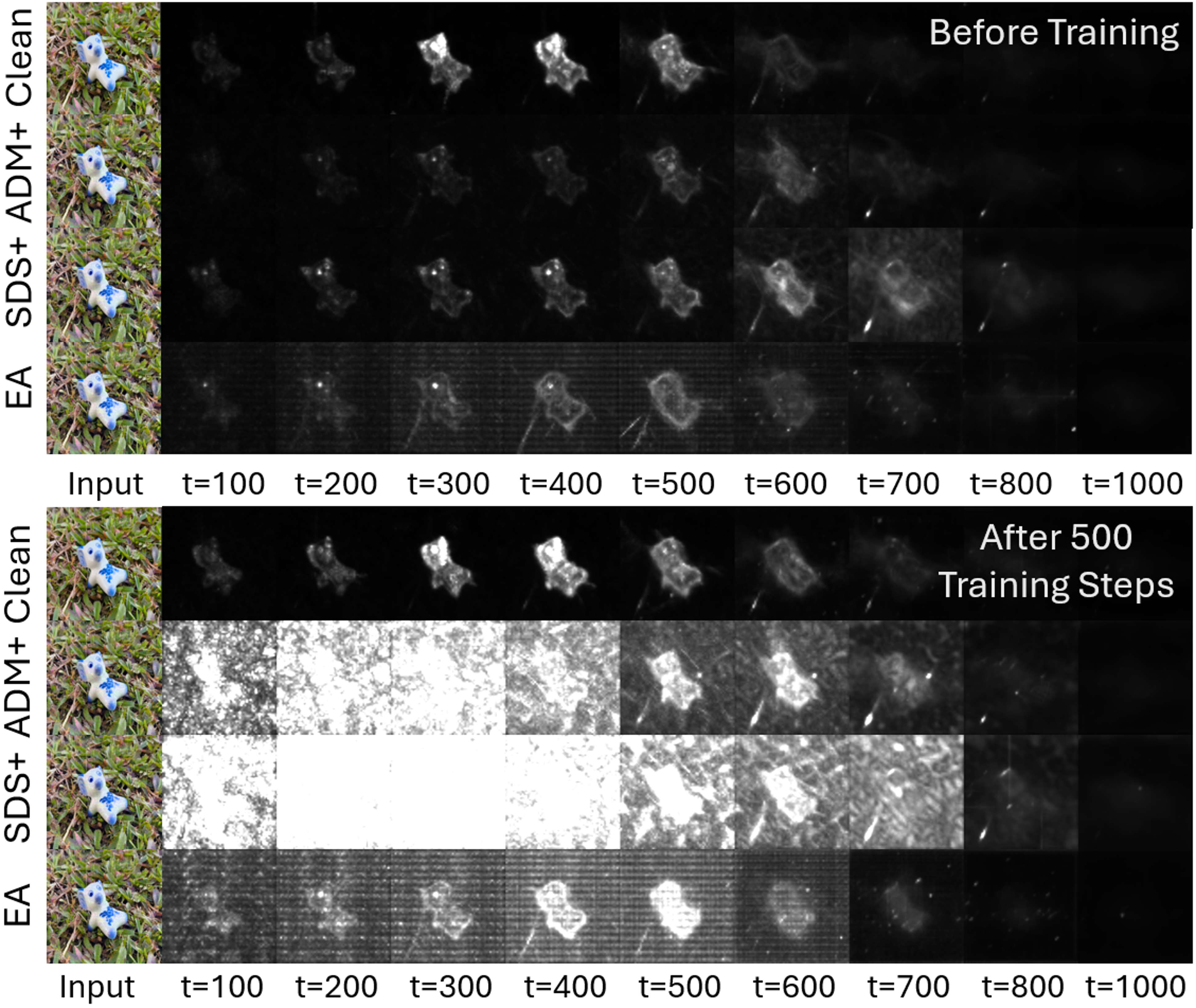}
  \caption{Comparison between SSMs of clean and AE samples at different noise timesteps (a) before TI training and (b) after 500 steps of TI training.}
  \label{fig:semantic_sensitivity_map_for_training}
  \vspace{-0.5cm}
\end{figure}

Using SSMs, we can investigate when and where adversarial signals attack the TI process. Fig. \ref{fig:semantic_sensitivity_map_for_training} shows the SSMs for a clean image and its corresponding AEs for various noising timesteps. The input images contain a novel concept not contained in the training dataset. The image prompt is ``a $R^*$ laying on top of a grass covered field". Fig. \ref{fig:semantic_sensitivity_map_for_training} shows the SSMs for the trainable text embedding corresponding to $R^*$ before and after TI training. We observe that before TI training, SSMs of the clean sample and AEs highlight the novel object in the range $t \in [300, 700]$. This indicates that the DM prior can already correlate unknown objects with unknown tokens. After TI training, the novel concept is more pronounced in SSMs of clean images, but the SSMs of AEs become extremely noisy, particularly at lower-middle timestep values. Notably, the attribution for AEs extends far outside the novel concept region. To explain this behavior, we investigate the temporal and spatial learning properties of TI in subsequent sections.

\subsection{Timestep Learning Bias}
\label{subsec:Timestep_Learning_Bias}

Fig. \ref{fig:semantic_sensitivity_map_for_training} shows that concept learning is focused on lower-middle timesteps ($t<700$) and that poison signals have the greatest impact at lower timesteps. To understand this phenomenon, we revisit the noising and training processes.

We begin by noting that all training objectives (i.e., model training in Eq. \ref{eqn:Noise_Pred}, TI in Eq. \ref{eqn:Textual_Inversion}, or AEs in Eq. \ref{eqn:Diffusion_Attack}) utilize the expected noise-prediction loss over a uniform $t$ distribution. The only differences between the objectives are the training datasets and the trainable parameters.

\begin{figure}[t]
  \centering
  \includegraphics[width=\linewidth,height=0.55\linewidth]{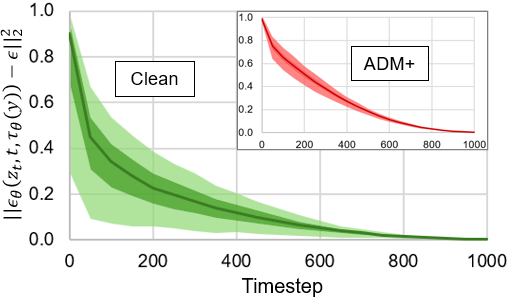}
  \caption{Loss distributions for Stable Diffusion 1.5 on $1000$ clean (green) and $1000$ ADM+ poisoned (red) LAION Aesthetic images at each of $21$ interspersed timesteps. The dark lines represent median loss values. The dark and light bands encompass 50\% and 95\%, respectively, of the loss distributions.}
  \label{fig:Loss_Distribution_Clean_M0+}
  \vspace{-0.5cm}
\end{figure}

Recalling that $\overline{\alpha}_{t}$ from Eq. \ref{eqn:Noising} monotonically decreases from $1$ at $t=0$ to $0$ at $t=T$, we can see that the noising process is interpolating between input $z_{0}$ and noise $\epsilon$. At $t=0$, $z_{t=0}=\sqrt{\overline{\alpha}_{0}}z_{0} + \sqrt{1-\overline{\alpha}_{0}}\epsilon=z_{0}$. Therefore, at $t=0$, the input to the noise-prediction model $\epsilon_{\theta}(z_0,0)$ contains no information about the sampled noise $\epsilon$. Since $\epsilon$ is distributed as $\mathcal{N}(\textbf{0},I)$, the optimal noise prediction to minimize the expected loss at $t=0$ is $\epsilon_{\theta}(z_0,0)=\textbf{0}$. Predictably, the loss at $t=0$ is a maximum because the noise-prediction model cannot accurately predict the ground-truth noise. Conversely, at $t=T$, $z_{t=T}=\sqrt{\overline{\alpha}_{T}}z_{0} + \sqrt{1-\overline{\alpha}_{T}}\epsilon=\epsilon$, and the noise-prediction model simply needs to return the input to minimize loss. At $t=T$, the loss approaches a minimum.

To validate our argument, we calculate the noise-prediction loss of Stable Diffusion 1.5 for $1000$ images sampled from the LAION Aesthetic \cite{2022_Schuhmann_LAION} dataset at $21$ equally-spaced timesteps ($21000$ samples total) and plot the resulting distributions in Fig. \ref{fig:Loss_Distribution_Clean_M0+}. As predicted, the expected loss decreases from a maximum at $t=0$ to $0$ at $t=T$. Applying ADM+ \cite{2023_Liang_AdvDM} poisoning to the images increases the loss values in the lower-middle timesteps (as expected by a loss-maximizing AE algorithm), but the losses at $t=0$ and at high timesteps remain unchanged.

\begin{figure}[t]
  \centering
  \includegraphics[width=\linewidth,height=0.55\linewidth]{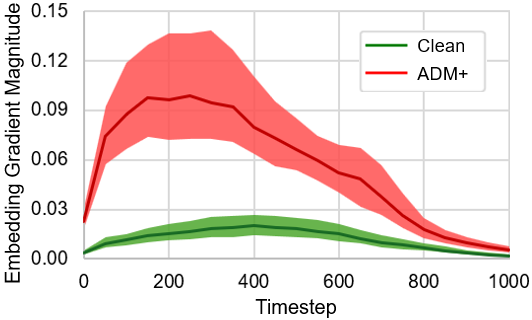}
  \caption{Trainable text embedding loss gradients after 1000 steps of TI training on clean and ADM+ poisoned NovelConcepts10 images. Results at each timestep are averaged across all concepts. The dark lines represent median gradient values. The dark bands encompass 50\% of the gradient distributions.}
  \label{fig:Clean_M0+_Grads}
  \vspace{-0.5cm}
\end{figure}

We must also consider how the noise-prediction objective affects gradients during TI training. Fig. \ref{fig:Clean_M0+_Grads} displays the average loss gradient magnitude with respect to the trainable text embedding at each timestep after $1000$ steps of TI training with Stable Diffusion 1.5. Curves are obtained by calculating $1000$ losses for each concept of the NovelConcepts10 dataset across $21$ equally-spaced timesteps and then averaging gradient magnitudes at each timestep across concepts. As expected, the graph shows that at high timesteps, gradients decay to $0$ as the loss distributions converge to $0$. At lower-middle timesteps, moderate loss allows the model to backpropagate real (or adversarial) signals to the text embedding, corresponding to high gradients. However, near $t=0$, high losses correspond to low gradients. Intuitively, this is because the noise-prediction model minimizes loss at $t=0$ by predicting near-$\textbf{0}$ noise regardless of the inputs, $z_{t}$ and $\tau_{\theta}(y^*)$. Since the prediction is independent of the inputs, no loss signal is backpropagated to the trainable text embedding. The loss gradient magnitudes in Fig. \ref{fig:Clean_M0+_Grads} correspond to the intensities of the sensitivity maps in Fig. \ref{fig:semantic_sensitivity_map_for_training}.

This behavior is particularly applicable to diffusion-based AEs and TI. All diffusion-based AEs seek to maximize the expected noise-prediction loss (Eq. \ref{eqn:Noise_Pred}), which includes an expectation over a uniform timestep range $\mathcal{U}(0,T)$. As the loss gradient is near $\textbf{0}$ at high timesteps, most of the adversarial signals will be learned at lower-middle timesteps. Likewise, TI applies an expectation over a uniform timestep range (Eq. \ref{eqn:Textual_Inversion}) thus focusing learning at lower-middle timesteps. Assuming that the adversarial signals contained within the training images are biased towards lower-middle timesteps, then we can bias the TI task towards high timesteps to avoid adversarial signals and learn clean concept features. 

In Appendix \ref{sec:appendix_proof}, we give a detailed derivation of the noise-prediction error along with additional plots for the loss and noise-prediction distributions. In Appendix \ref{sec:appendix_gradients}, we show the resulting effect on loss gradients throughout TI training and across multiple poisons.

\subsection{Spatial Learning Bias}
\label{subsec:Spatial_Learning_Bias}

As observed in Sec. \ref{subsec:Semantic_Sensitivity_Map}, AEs disrupt TI by ``distracting" the optimization process and expanding the spatial distribution of the trainable token attribution beyond the novel concept region. To understand this phenomenon, we revisit the adversarial objectives in Eq. \ref{eqn:Diffusion_Attack} and Eq. \ref{eqn:Encoder_Attack}.

We begin by noting that both adversarial objectives involve an $l^2$ norm which includes a sum over all spatial coordinates (height, width, channels) of the error tensor. This summation does not apply preferential weight to any spatial region. Furthermore, the $l^2$ vector norm is convex and non-decreasing, implying that it increases as the magnitudes of individual elements increase. This encourages all spatial elements to contribute to optimizing the $l^2$ objective. This incentive combines with the adversarial perturbation constraint $||\delta||_b$, which is commonly implemented with the $l^\infty$ norm. We note that $||\delta||_\infty$ is non-increasing for all infinitesimal changes in perturbation elements $\delta_i$ where $|\delta_i|<||\delta||_\infty$; this permits increases in most perturbation elements without affecting the constraint. In summary, the $l^2$ norm objectives in Eq. \ref{eqn:Diffusion_Attack} and Eq. \ref{eqn:Encoder_Attack} encourage all elements of the error terms to contribute and the $l^\infty$ constraint on $\delta$ encourages all elements of $\delta$ to contribute maximally. Practically, this implies that adversarial perturbations maximally infect every region of a poisoned image.

\begin{figure}[t]
  \centering
  \includegraphics[width=\linewidth]{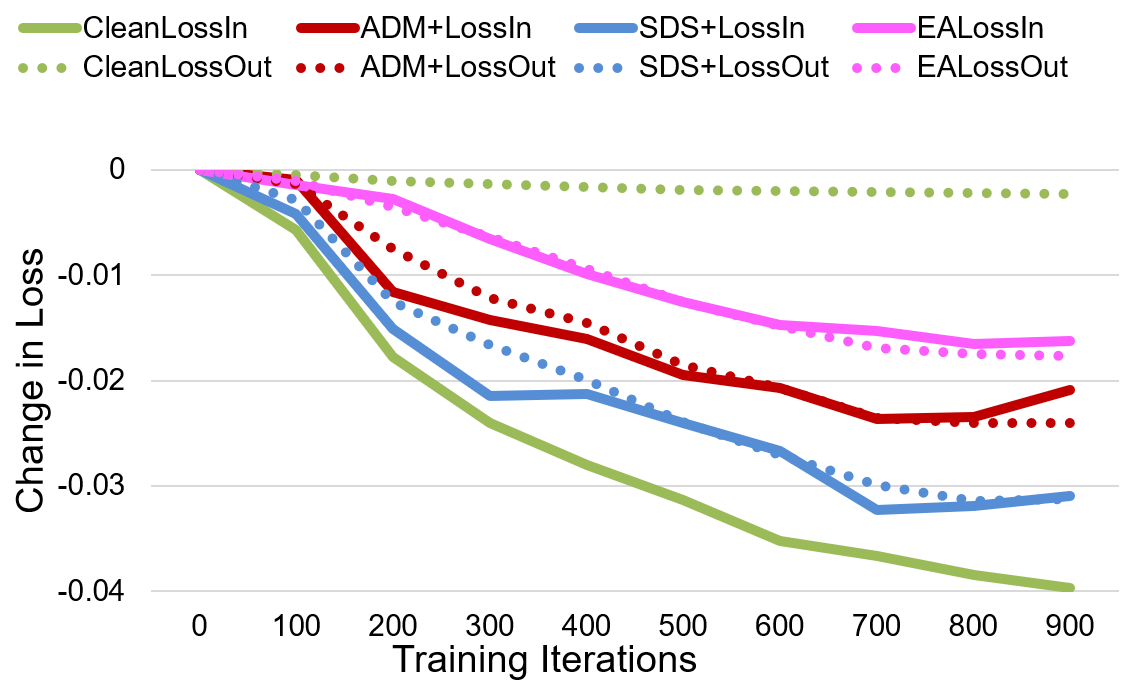}
  \caption{Loss inside (``LossIn") and outside (``LossOut") novel concept regions during TI training on clean, ADM+, SDS+, and EA data, averaged across all concepts in NovelConcepts10.}
  \label{fig:loss_reduced}
\end{figure}

To empirically verify this phenomenon, we study the losses attributed to novel concept regions for clean images and AEs during TI training. We apply TI to clean, ADM+, SDS+, and EA versions of the NovelConcepts10 dataset. We track the loss contributions from inside and outside the novel concept regions throughout training. We first average losses across their contributing pixels and then average the results across all concepts. The results in Fig. \ref{fig:loss_reduced} indicate that TI for clean images can effectively minimize the loss inside the novel concept regions and ignores signals outside these regions. This implies that TI effectively learns new concepts and ignores background information. However, when trained on AEs, TI losses inside and outside of the concept regions are reduced by the same proportion. This indicates that AEs mislead TI to learn equally from all regions and ignore the semantic significance of the novel concept. This aligns with the observation from Fig. \ref{fig:semantic_sensitivity_map_for_training} that all regions of an adversarial input are attributed to the trainable token.

\subsection{JPEG as Poison Defense}
\label{subsec:JPEG_Analysis}

The adversarial signals injected into images by poisoning techniques tend towards higher frequencies in pixel space. Consequently, JPEG compression, which relies on discrete cosine transforms to identify and remove high-frequency signals, has been noted as an effective poison defense \cite{2023_Liu_ISS}. We investigate the mechanism by which JPEG eliminates adversarial signals and identify two crucial behaviors: (1) JPEG compression converts bimodal poison noise into unimodal Gaussian-like noise in pixel space and (2) JPEG centralizes the frequency spectrum in latent space, forcing poisoned latent frequencies towards those of clean latents. 

\begin{figure}[t]
  \centering
  \includegraphics[width=\linewidth,height=0.45\linewidth]{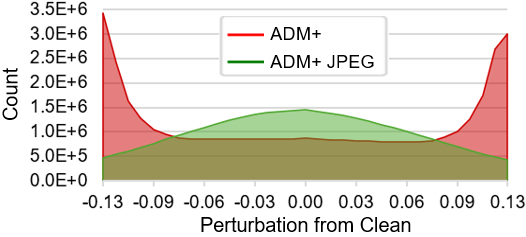}
  \caption{Histograms of pixel-space perturbations (relative to clean images) for ADM+ poisoned images without (red) and with (green) JPEG compression.}
  \label{fig:JPEG_Hist_ADM}
\end{figure}

We empirically verify (1) in Fig. \ref{fig:JPEG_Hist_ADM} by examining the distribution of poison perturbations (relative to clean images) with and without applying JPEG compression. We use a JPEG quality of 25 and aggregate results across all 50 images in the NovelConcepts10 dataset. The bimodal distribution of poison perturbations aligns with loss objective and constraint discussion from Sec. \ref{subsec:Spatial_Learning_Bias}. After JPEG compression, the poison perturbations follow a unimodal Gaussian-like distribution. We note in Fig. \ref{fig:JPEG_Hist_All} of Appendix \ref{sec:appendix_JPEG_compression} that this behavior holds for all poisons studied. Intuitively, the carefully-optimized poison signals have been replaced by a distribution of random errors due to JPEG compression.

Due to space limitations, we move empirical verification of (2) and subsequent analysis to Appendix \ref{sec:appendix_JPEG_compression}. We note that the effect of (2) is that JPEG-compressed poisoned images force poison latents to conform to the power spectra of natural image latents, as would be expected by LDMs.

\subsection{Safe-Zone Training}
\label{subsec:SZT_Method}

In the above analyses, we have demonstrated when and where AEs poison TI. Combining takeaways from our analyses, we propose Safe-Zone Training (SZT) to mitigate the influence of adversarial signals in poisoned images. Applying lessons from Sec. \ref{subsec:JPEG_Analysis}, we first apply JPEG compression to obfuscate the adversarial signals in the pixel domain and centralize the power spectra in the latent domain. Applying the conclusions from Sec. \ref{subsec:Timestep_Learning_Bias}, we restrict training to high timesteps in order to avoid adversarial signals at lower-middle timesteps. Finally, SZT incorporates the behaviors noted in Sec. \ref{subsec:Spatial_Learning_Bias} and utilizes a binary loss mask to allow only signals from the novel concept region to backpropagate to the trainable text embedding. In summary, SZT is a simple algorithm that involves applying JPEG compression, adjusting the timestep sampling range, and inserting a mask during loss calculation. Crucially, SZT does not require additional networks to regenerate or purify poison data. The SZT objective can be formulated as
\begin{align}
    \label{eqn:SZT}
    & \argmin_{\theta_{R^*}} E_{x'_{j}, t, y^*, \epsilon} \left[||(\epsilon_{\theta}(z'_{t},t,\tau_{\theta}(y^*)) - \epsilon) \odot M_{z}||_{2}^2 \right] \\
    & \text{s.t.} \ x'_{j} = JPEG(x'_{p}), \ t \sim \mathcal{U}(t_{th},T), \ M_{z} = Resize(M_x), \notag
\end{align}
where $x'_j$ is the JPEG-compressed version of poisoned image $x'_{p}=x_{p}+\delta$, $z'_{t}=\sqrt{\overline{\alpha}_{t}}\mathcal{E}(x'_{j}) + \sqrt{1-\overline{\alpha}_{t}}\epsilon$, timestep $t$ is uniformly sampled from the range above threshold $t_{th}$, $M_{z}$ is the latent-space mask resized from the pixel-space mask of the novel concept region $M_{x}$, and $\odot$ denotes element-wise multiplication.
\section{Experiments}
\label{sec:experiments}

\subsection{Experiment Settings}

\subsubsection{Datasets}
\label{subsubsec:5_Experiments_Experiment-Settings_Datasets}

We test the ability of TI to learn new concepts from poisoned images. To ensure that the LDM is learning new concepts not contained within the prior, we collected NovelConcepts10, a dataset of five images for each of 10 distinct objects (50 total images). An extended dataset description and example images are given in Appendix \ref{sec:appendix_novel_concepts}. We also evaluate on CustomConcept101 \cite{2023_Kumari_CustomDiffusion}, which was curated to study personalization tasks. For each image in a dataset, we create a binary mask of the novel concept region using the Segment Anything Model \cite{2023_Kirillov_SegmentAnything}. We utilize six data poisons: ADM+ \cite{2023_Liang_AdvDM}, ADM-, SDS+ \cite{2024_Xue_SDS}, SDS- \cite{2024_Xue_SDS}, EA \cite{2023_Salman_Photoguard}, and DA \cite{2023_Salman_Photoguard}. We place poisoning details in Appendix \ref{sec:appendix_poisons}. 

\subsubsection{Model and Training}
\label{subsubsec:Model_Training}

We conduct all experiments with Stable Diffusion v1.5 \cite{2022_Rombach_LDM}. We utilize a modified Diffusers library from HuggingFace \cite{2022_Von-Platen_Diffusers} to perform TI. For each concept of a dataset, we run TI \cite{2022_Gal_TextInv} with a default learning rate of $5e^{-4}$ (no decay, no warmup) for $5000$ training steps and a batch size of $1$. For CustomConcept101, we utilize $2500$ training steps without noticeable loss of performance. For each learned concept, we generate five images for each of the $24$ object prompts from the DreamBooth pipeline \cite{2023_Ruiz_DreamBooth} ($120$ images total). 

\subsubsection{Evaluation Metrics}

For each learned concept, we measure the DINOv2 encoding similarity \cite{2024_Oquab_DINOv2} and FID score \cite{2018_Heusel_FID} between training images and the images generated after TI. For poisoned datasets, we calculate metrics between generated images and clean versions of the training images. We also measure prompt fidelity with CLIP Score \cite{2021_Radford_CLIP, 2022_Hessel_ClipScore} between the generation prompt and the generated images. We report metrics first averaged across all generated images per concept and then across all concepts in a dataset.

\subsection{Results and Analysis}

\subsubsection{Personalization Settings Ablations}

We begin by ablating training hyperparameters for TI across clean, ADM+, ADM-, and EA versions of NovelConcepts10 for learning rate in ($5e^{-5}$, $5e^{-4}$, $5e^{-3}$), learning rate schedule in (constant, linear decay, cosine decay), and training steps in ($1000$, $5000$, $10000$). Results and further discussion are given in Appendix \ref{sec:appendix_hyperparams}. In summary, we find that maintaining default settings (constant $5e^{-4}$ learning rate with $5000$ training steps) is sufficient to learn novel concepts without overfitting to the training images.


\subsubsection{Timestep Range Ablation}

Based on findings about timestep learning bias in Sec. \ref{subsec:Timestep_Learning_Bias}, we ablate methods to restrict timesteps to high values during TI training. We consider multiple sampling methods to favor high timesteps: thresholding, power distributions, and $tanh$ distributions. Fig. \ref{fig:Prob_Curves} in Appendix \ref{sec:appendix_trange} displays the probability curves for the power and $tanh$ distributions. We ablate $t$-sampling methods for clean, ADM+, \mbox{ADM-,} and EA versions of NovelConcepts10. Full results are given in Appendix \ref{sec:appendix_trange}. We find that simple high-thresholding works best, improving DINOv2 similarity for ADM+ data from $0.10$ ($t \sim \mathcal{U}(0,1000))$ to $0.21$ for $t \ge 600$ and $0.28$ for $t \ge 700$. Using $t \ge 600$ also maintains or improves scores for clean and EA datasets. In our subsequent proposed defenses, we mainly utilize $t \ge 600$, denoted ``T600".

\subsubsection{Masking Ablation}
\label{subsubsec:Masking_Ablation}

To verify the effectiveness of loss masking (LM), we compare it to various masking strategies: input image masking (IM), latent ($z_t$) masking (ZM), and masking both input image and loss (LIM). We ablate the masking method on all poisoned versions of NovelConcepts10. Results in Appendix \ref{sec:appendix_formulation_Masking} demonstrate that LM outperforms all other masking strategies across all poisoning methods, except EA. We hypothesize that applying masks directly to the input image or latent $z_t$ may destroy useful information before the features are processed by the U-Net. Intuitively, masked inputs lack contextual information about the relationship between the novel concept and the background, distorting the U-Net predictions. By applying the mask to the final loss, we preserve background information and permit a more coherent understanding of the scene.

In Appendix \ref{sec:appendix_maskdilate}, we experiment with dilating the mask size to incorporate additional context in loss backpropagation. We find that relatively small amounts of mask dilation (i.e., $16$ pixels for $512$x$512$ masks) can improve LM for TI.

\subsubsection{Safe-Zone Training Performance}
\label{subsubsec:SZT_Performance}

\begin{table*}
  \centering
  \caption{CustomConcept101 DINOv2 Similarity for various poison defenses.}
  \begin{tabular}{l|c|cccc|ccccc}
    \toprule
    Defense $\rightarrow$ & Nominal & Regen & PDMPure & AdvClean & JPEG & T600 & LM & JPEG & JPEG & SZT \\
    Poison $\downarrow$ & & \cite{2024_Zhao_Regeneration} & \cite{2024_Xue_PDMPure} & \cite{2025_Shidoto_AdverseCleaner} & & & & +T600 & +LM & \\
    \midrule
    Clean & 0.46 & 0.46 & 0.25 & 0.45 & 0.45 & \textbf{0.50} & 0.47 & 0.47 & 0.46 & \underline{0.48} \\
    \midrule
    ADM+\cite{2023_Liang_AdvDM} & 0.10 & 0.17 & 0.23 & 0.15 & 0.36 & 0.25 & 0.30 & 0.41 & \underline{0.44} & \textbf{0.45} \\
    ADM- & 0.33 & 0.25 & 0.23 & 0.35 & 0.41 & 0.27 & 0.43 & \underline{0.45} & \underline{0.45} & \textbf{0.46} \\
    SDS+\cite{2024_Xue_SDS} & 0.10 & 0.16 & 0.25 & 0.16 & 0.34 & 0.23 & 0.32 & 0.40 & \textbf{0.45} & \underline{0.44} \\
    SDS-\cite{2024_Xue_SDS} & 0.31 & 0.22 & 0.23 & 0.35 & 0.40 & 0.23 & 0.42 & \underline{0.45} & \underline{0.45} & \textbf{0.47} \\
    EA\cite{2023_Salman_Photoguard} & 0.10 & 0.09 & 0.24 & 0.19 & 0.33 & 0.12 & 0.29 & \underline{0.44} & \underline{0.44} & \textbf{0.46} \\
    DA\cite{2023_Salman_Photoguard} & 0.29 & 0.25 & 0.27 & 0.35 & 0.39 & 0.22 & 0.42 & 0.44 & \underline{0.46} & \textbf{0.47} \\
    \midrule
    Psn Avg & 0.21 & 0.19 & 0.24 & 0.26 & 0.37 & 0.22 & 0.36 & 0.43 & \underline{0.45} & \textbf{0.46} \\
    \bottomrule
  \end{tabular}
  \label{tab:CC101_Baseline_DINOv2}
\end{table*}

\begin{figure*}[h]
  \centering
  \includegraphics[width=\linewidth]{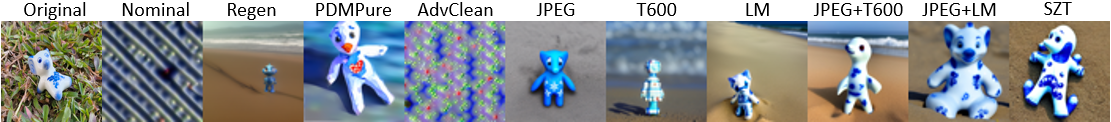}
  \caption{Generated images after TI on NovelConcepts10 BlueElephant poisoned by ADM+. Generation prompt: ``a $R*$ on the beach"}
  \label{fig:GenImgs_Main}
\end{figure*}
\vspace{-4pt}


Our implementation of SZT is highly customizable, empowering the user to control JPEG compression quality, timestep restriction method/shape, and loss mask size/shape. In our ablations, we find that combining JPEG compression (quality$=25$), timestep restriction $t\ge500$, and LM with $16$ pixels of mask dilation performs best. We defer further discussion on SZT configurations to Appendix \ref{sec:appendix_maskdilate} and encourage users to explore additional settings.

We demonstrate the robustness of SZT across poisoned versions of the CustomConcept101 dataset. Table \ref{tab:CC101_Baseline_DINOv2} compares the DINOv2 similarity performance of SZT and its novel components with established defenses, including Regen \cite{2024_Zhao_Regeneration}, PDMPure \cite{2024_Xue_PDMPure}, AdvClean \cite{2025_Shidoto_AdverseCleaner}, and JPEG compression; we apply Regen default settings with Stable Diffusion 1.5, utilize published repositories for PDMPure and AdvClean, and set JPEG quality to $25$. 

JPEG is a moderately-strong defense, outperforming established methods like Regen and PDMPure on multiple poisons. T600, despite requiring only one parameter change and no extra computation overhead, modestly improves performance against adversarial poisons like ADM+ and SDS+ and outperforms some established defenses on these poisons. LM is a strong standalone defense, improving the generative quality for every poison. Adding JPEG compression as a preprocessing step for T600 or LM further improves performance across all poisons. In particular, JPEG+T600 is an incredibly efficient defense, delivering competitive image quality across all poisons without additional networks or masking. SZT is consistently the strongest defense, achieving state-of-the-art image quality across almost all poisons. Notably, SZT performance on all \underline{poison} data (average of $0.46$) matches the quality achieved by nominal TI on \underline{clean} data, effectively nullifying all poisons. We also report FID and CLIP Score for CustomConcept101 in Appendix \ref{sec:appendix_CC101_baselines}. We perform the same study on NovelConcepts10 in Appendix \ref{sec:appendix_NC10_baselines} and find that the trends are largely similar.

We include studies on additional LDM architectures and parameterizations in Appendix \ref{sec:appendix_Other_Models} and additional personalization methods in Appendix \ref{sec:appendix_Other_Personalization}. Across all models and methods, SZT and its ablations are the strongest defenses.

For a qualitative comparison, we display images generated after TI training on the BlueElephant concept poisoned by ADM+ in Fig. \ref{fig:GenImgs_Main}. Existing defenses (Regen, PDMPure, and AdvClean) can occasionally improve generated image quality, resulting in generated images that contain a semblance of the desired concept. Images generated after JPEG compression display some faithfulness to the desired concept. T600 also presents as a moderately effective defense, often removing poison noise but occasionally failing to capture the desired concept. LM alone tends to learn the correct concept shape but not the correct texture or color, as LM does not attempt to mitigate adversarial signals within the masked region. JPEG+LM improves on this weakness. JPEG+T600 strikes a good balance between accurately learning the desired concept and displaying prompt context. SZT is able to accurately learn the desired concept shape and texture. We note that defenses utilizing the LM operation (i.e., LM, JPEG+LM, and SZT) tend to hyperfocus on the concept, generating images that are dominated by the concept. This behavior is reduced for SZT, which uses dilated masks to incorporate some background information from training images. We include further qualitative results for additional poisons and concepts from CustomConcept101 and NovelConcepts10 in Appendix \ref{sec:appendix_gen_imgs}. 

\section{Conclusion}

The SZT defense method is derived from systematic analyses of the underlying mechanisms of DMs as well as observations on the learning tendencies of TI. Despite the simplicity of SZT, it is an effective defense method that performs beyond other established defenses and nullifies many poisons. We hope that our work exposes the vulnerabilities of existing poisons and spurs further research on robust poisons for DMs.

\clearpage
\section*{Acknowledgments}



This research is supported by the National Research Foundation, Singapore and Infocomm Media Development Authority under its Trust Tech Funding Initiative, and ROSE @ NTU. Any opinions, findings and conclusions or recommendations expressed in this material are those of the author(s) and do not reflect the views of National Research Foundation, Singapore and Infocomm Media Development Authority, or ROSE @ NTU.
{
    \small
    \bibliographystyle{ieeenat_fullname}
    \bibliography{main}
}

\clearpage
\appendix{}
\label{sec:appendix}


\section{Additional Semantic Sensitivity Maps}

\begin{figure}[t]
  \centering
  \includegraphics[width=\linewidth]{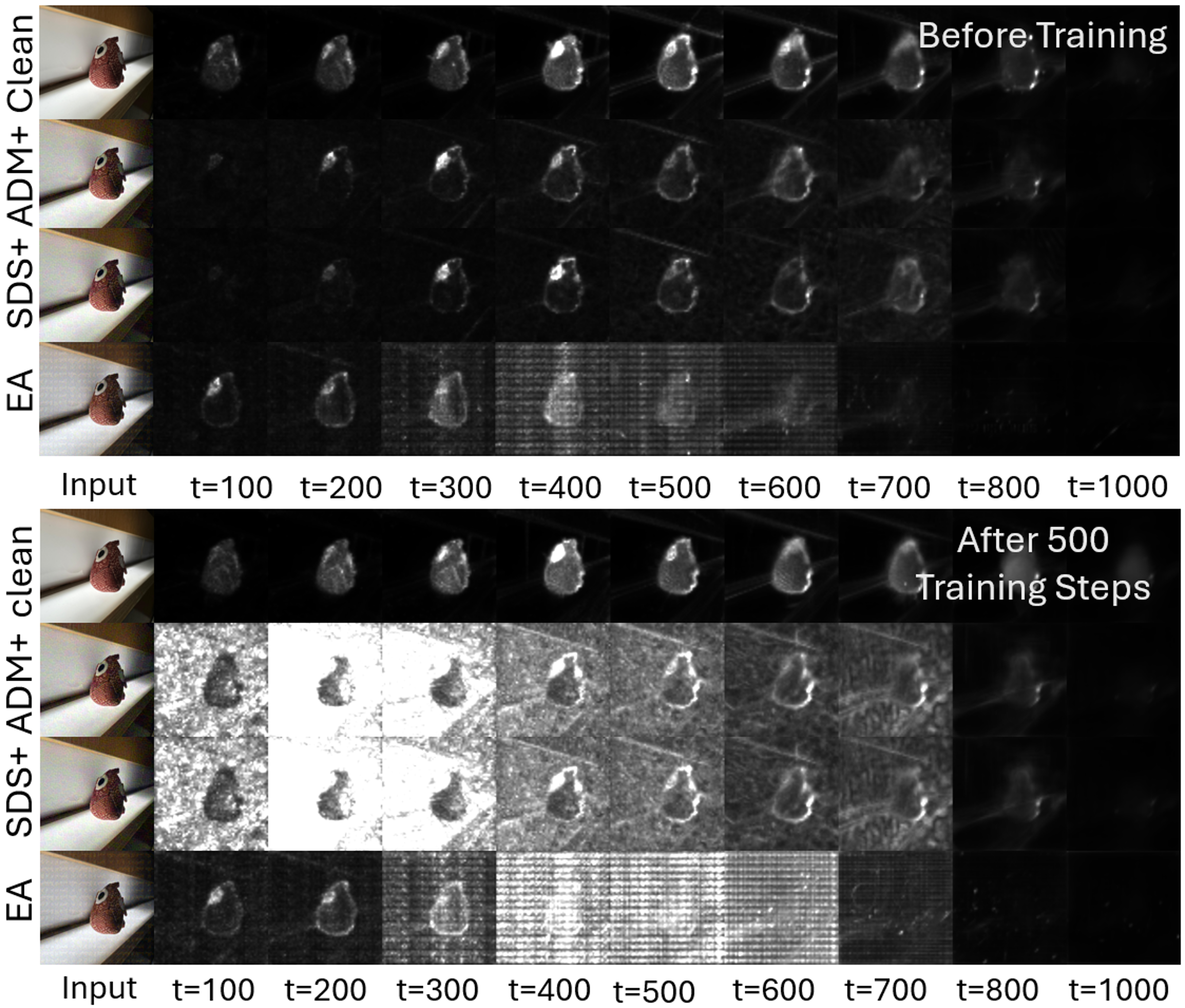}
  \caption{Comparison between SSMs of clean and AE samples at different noise timesteps (a) before TI training and (b) after 500 steps of TI training.}
  \label{fig:more_SSM_comparison}
\end{figure}

We include an additional study of SSMs for the FishDoll concept from NovelConcepts10 in Fig. \ref{fig:more_SSM_comparison}. Trends here generally reflect those noted in Sec. \ref{subsec:Semantic_Sensitivity_Map}.

\section{Noise-Prediction Distribution}
\label{sec:appendix_proof}

\begin{figure}[t]
  \centering
  \includegraphics[width=\linewidth,height=1.05\linewidth]{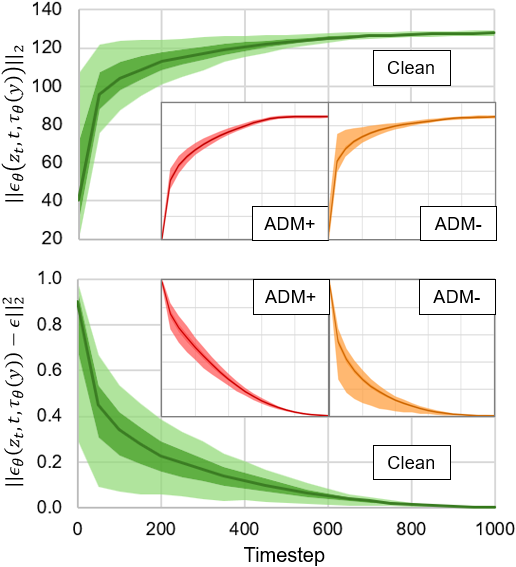}
  \caption{Noise prediction magnitude (top) and loss (bottom) distributions for Stable Diffusion 1.5 on 1000 LAION Aesthetic images at each of 21 interspersed timesteps.}
  \label{fig:L2_Loss_Combined_Clean_M0+_M0-}
\end{figure}

We assume that the distribution of LDM latents $z_{0}$ follows some Gaussian $\mathcal{N}(\mu_{z},\Sigma_{z})$.\footnote{We note a distinction between distributions $p(z_0|x)$ and $p(z_0)$. In LDMs where the encoder is from a VAE \cite{2022_Kingma_VAE, 2022_Rombach_LDM}, $p(z_0|x)$ is Gaussian by definition, as $z_{0}$ is randomly sampled from a multivariate Gaussian that describes the encoding space (i.e., $\mathcal{E}(x)=[\mu_{z},\Sigma_{z}]$). However, across many samples $x$, $p(z_0)$ is not necessarily Gaussian. Our analysis in this section is exactly correct in the single-image case ($p(z_0|x)$), but its correctness in the multi-image case ($p(z_0)$) hinges on the correctness of the Gaussian assumption for $p(z_0)$. In a brief experiment (not shown), we sample $z_0$ for $1000$ LAION Aesthetic images and find that the distributions of $z_0$ dimensions are generally unimodal and akin to bell-curves, but they are decidedly not Gaussian. Even so, the empirical results for the multi-image case in Fig \ref{fig:L2_Loss_Combined_Clean_M0+_M0-} conform to the behavior expected by our derivation in this section, suggesting that the Gaussian assumption for $p(z_0)$ is permissible.} The ground truth noise $\epsilon$ is sampled from a standard Gaussian $\mathcal{N}(\textbf{0},I)$. For some timestep $t$, we calculate $z_{t}$ following Eq. \ref{eqn:Noising}. The distribution of $z_{t}$ can be described as the Gaussian achieved by scaling and combining the $z_{0}$ and $\epsilon$ distributions.
\begin{equation}
    \label{eqn:zt_Distribution}
    z_{t} \sim \mathcal{N}(\sqrt{\overline{\alpha}_{t}}\mu_{z},\Sigma_{zt}),
\end{equation}
where $\Sigma_{zt}=\overline{\alpha}_{t}\Sigma_{z}+(1-\overline{\alpha}_{t})I$. We assume that $\Sigma_{zt}$ is invertible. The joint distribution of $\epsilon$ and $z_{t}$ is given by
\begin{equation}
    \label{eqn:Joint_Distribution}
    p(\epsilon,z_{t}|t) = \mathcal{N} \left(
        \begin{bmatrix}
            0\\
            \sqrt{\overline{\alpha}_{t}}\mu_{z}
        \end{bmatrix}, \\
        \begin{bmatrix}
            I & \sqrt{1-\overline{\alpha}_{t}}I \\
            \sqrt{1-\overline{\alpha}_{t}}I & \Sigma_{zt}
        \end{bmatrix}
    \right ),
\end{equation}
where we apply the linearity property of covariance ($cov(A,A+B)=cov(A,A)+cov(A,B)$) and the independence between $z_{0}$ and $\epsilon$ to derive the off-diagonal covariance matrices. We seek to predict the distribution of noise when given the same inputs as a noise-prediction DM. We use the definition of a conditional distribution, $p(x|y)=p(x,y)/p(y)$ to derive the distribution of $\epsilon$ conditioned upon inputs $z_{t}$ and $t$. The conditional distribution can be derived in closed form from the joint distribution \cite{2025_Wikipedia_MultivarGaussian}.
\begin{equation}
    \label{eqn:Cond_Distribution}
    \begin{split}
        p(\epsilon|z_{t},t)
        = \mathcal{N}( & \sqrt{1-\overline{\alpha}_{t}}(\Sigma_{zt})^{-1}(z_{t}-\sqrt{\overline{\alpha}_{t}}\mu_{x}), \\
        & I-(1-\overline{\alpha}_{t})(\Sigma_{zt})^{-1}),
    \end{split}
\end{equation}

Though initially complicated, this conditional distribution simplifies beautifully in the limits as $t \rightarrow 0$ and $t \rightarrow T$. As $t \rightarrow 0$, $\overline{\alpha}_{t} \rightarrow 1$, so the conditional distribution approaches $\mathcal{N}(\textbf{0},I)$. This is sensible because $z_{t} \rightarrow z_{0}$, which contains no information about the ground-truth noise, so the noise distribution $\epsilon=\mathcal{N}(\textbf{0},I)$ is independent of $z_{0}$. As $t \rightarrow T$, $\overline{\alpha}_{t} \rightarrow 0$, so the distribution approaches $\mathcal{N}(z_{T},\textbf{0})$. This is sensible as $z_{t} \rightarrow z_{T}=\epsilon$.

\begin{figure*}[t]
  \centering
  \includegraphics[width=\linewidth,height=0.22\linewidth]{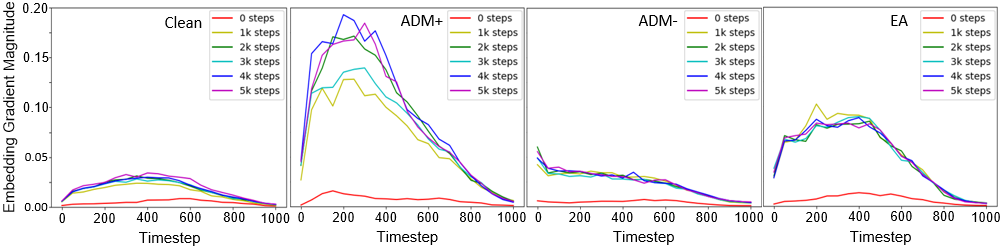}
  \caption{Plots of text embedding loss gradient magnitude as a function of timestep measured throughout training. Values at each timestep are averaged across concepts in the NovelConcepts10 dataset.}
  \label{fig:Embedding_Gradients}
\end{figure*}

This derivation can be extended to the distribution of the noise-prediction error (Eq. \ref{eqn:Noise_Pred}). As both components are Gaussian, the noise-prediction error $p(\epsilon|z_{t},t)-\epsilon$ is also Gaussian and the covariance of Eq. \ref{eqn:Cond_Distribution} is shifted by $I$. Notably, the noise prediction error has maximum variance at $t=0$ which monotonically decreases as $t$ increases.
\begin{equation}
    \label{eqn:NPE_Distribution}
    \begin{split}
        p(\epsilon|z_{t},t) & - \mathcal{N}(\textbf{0},I) \\
        = \mathcal{N}( & \sqrt{1-\overline{\alpha}_{t}}(\Sigma_{zt})^{-1}(z_{t}-\sqrt{\overline{\alpha}_{t}}\mu_{x}), \\
        & 2I-(1-\overline{\alpha}_{t})(\Sigma_{zt})^{-1}),
    \end{split}
\end{equation}

\begin{figure*}[t]
  \centering
  \includegraphics[width=\linewidth]{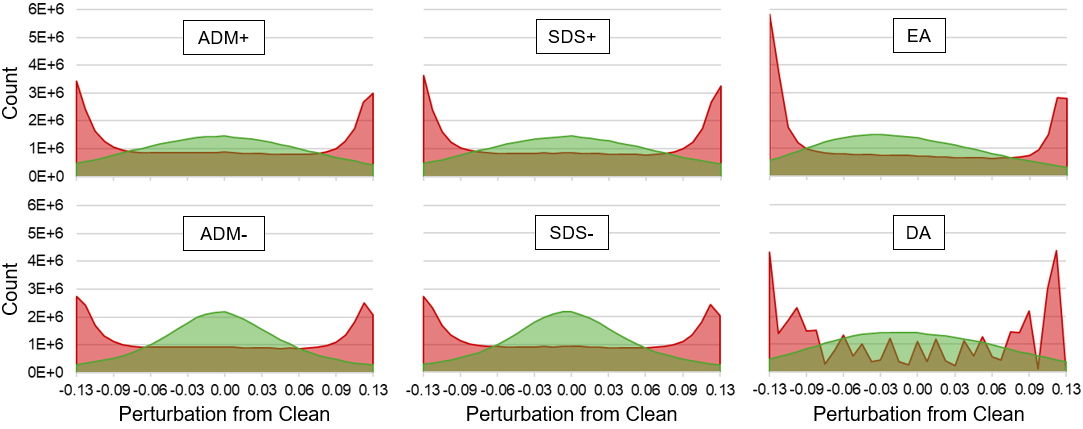}
  \caption{Histograms of pixel-space perturbations (relative to clean images) for poisoned NovelConcept10 images without (red) and with (green) JPEG compression. Input images are shifted and scaled to $[-1, 1]$}
  \label{fig:JPEG_Hist_All}
\end{figure*}

Fig. \ref{fig:L2_Loss_Combined_Clean_M0+_M0-} is an expanded version of Fig. \ref{fig:Loss_Distribution_Clean_M0+} that includes plots for the magnitudes of the predicted noise. As discussed in Sec. \ref{subsec:Timestep_Learning_Bias}, the model predicts noise near \textbf{$0$} at $t=0$ and perfectly predicts noise at $t=T$. The magnitude of the predicted noise grows from near \textbf{$0$} at $t=0$ to a maximum of $128$ at $t=T$ (which aligns with $E[||\epsilon||_{2}]=\sqrt{tr(I)}=128$ when $\epsilon$ is a 4x64x64 vector as in Stable Diffusion). The median loss monotonically decreases from a maximum at $t=0$ to $0$ at $t=T$. Additionally, the width of the loss distribution at each timestep indicates that the variance of the noise-prediction error is indeed a maximum at $t=0$ and diminishes to $0$ at $t=T$.

The noise-prediction DM always seeks to minimize loss by maximizing the conditional probability of the noise (Eq. \ref{eqn:NPE_Distribution}). At $t=0$, it will return the mean of $\epsilon \sim \mathcal{N}(\textbf{0},I)$. At high timesteps, it will increasingly predict its own input. At middle timesteps, it will actually learn to distinguish noise from image. Therefore, all useful learning will occur at lower-middle timesteps. 

Lastly, we note that this behavior is reversed (but still present) for input-prediction models (i.e., $z_{0,\theta}(\cdot)$). An input-prediction model will return its input at $t=0$ and the average image (i.e., a plain gray image) at $t=T$.


\section{Textual Inversion Gradients}
\label{sec:appendix_gradients}

In Fig. \ref{fig:Embedding_Gradients} we display the $l^{2}$ magnitude of the loss gradient on the trainable text embedding as a function of timestep $t$ throughout TI training. For each concept in the NovelConcepts10 dataset, we extract checkpoints every 1000 steps during TI training and calculate the expected gradient magnitude for each timestep then average the magnitudes across concepts. Aligning with our findings from SSMs (Sec. \ref{subsec:Semantic_Sensitivity_Map}) and timestep learning bias (Sec. \ref{subsec:Timestep_Learning_Bias}), all gradient magnitudes are biased towards lower-middle timesteps, with near-$0$ gradient magnitudes at high timesteps for all datasets. The ADM- poison, which seeks to minimize DM loss, mitigates the low-timestep bias. The EA poison, which optimizes poisons on the LDM encoder and avoids the DM entirely during optimization, still permits a bias towards lower timesteps during TI training. This is likely because EA focuses on perturbing $z_0$, thereby adversarially affecting high signal-to-noise $z_t$ latents (i.e., those at lower-middle timesteps).

We note that the dropoff in gradient at very low timesteps for Clean, ADM+, and EA samples is due to the tendency of the noise-prediction model to predict \textbf{0} at $t=0$. Although the loss at this point is maximal (as shown in Fig. \ref{fig:L2_Loss_Combined_Clean_M0+_M0-} (bottom)), the conditional distribution of $\epsilon$ (from Eq. \ref{eqn:Cond_Distribution}) at $t=0$ is independent of $z_{t}$, as noted in Sec. \ref{sec:appendix_proof}. Therefore, there is no loss gradient with respect to the inputs at $t=0$. In summary, learning is generally minimal at high timesteps and at $t=0$, regardless of whether the model is learning a true concept or an adversarial signal.

\section{JPEG Compression Analysis}
\label{sec:appendix_JPEG_compression}

In Fig. \ref{fig:JPEG_Hist_All}, we extend the analysis performed for Fig. \ref{fig:JPEG_Hist_ADM} in Sec. \ref{subsec:JPEG_Analysis} to all poisons. Across all poisons, we find that JPEG compression has the same effect: the poison perturbation distribution is converted from a bimodal distribution with modes that skew towards the perturbation limits ($\pm \kappa$) to a bell-curve distribution centered at or near $0$.

\begin{figure}[t]
  \centering
  \includegraphics[width=\linewidth]{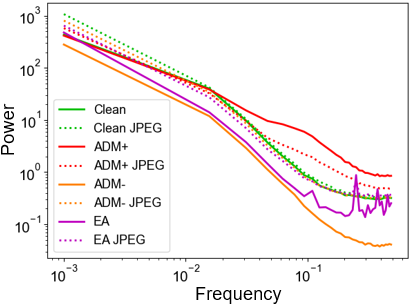}
  \caption{Radially averaged power spectrum density curves for mean latents with and without JPEG compression. Curves are averaged across all images in NovelConcepts10.}
  \label{fig:JPEG_Lat_RAPSD}
\end{figure}

To verify claim (2) from Sec \ref{subsec:JPEG_Analysis}, we also investigate the impact of JPEG compression on the LDM latent space. Fig. \ref{fig:JPEG_Lat_RAPSD} displays the radially averaged power spectrum density curves for the mean latent encodings of Clean, ADM+, ADM-, and EA images with and without JPEG compression, averaged across all 50 images in the NovelConcepts10 dataset. Before JPEG compression, the power curve for ADM+ latents is consistently higher than that of clean latents at high frequencies, whereas the powers for ADM- and EA latents are consistently lower. However, after JPEG compression (in pixel space), the power spectra of all images are centralized, lying closer to the power spectra of clean images. This result suggests that JPEG compression tends to ``standardize" poisoned images, forcing their latents to conform to the power spectra expected by LDMs.

\section{NovelConcepts10 Dataset}
\label{sec:appendix_novel_concepts}

\begin{figure}[t]
  \centering
  \includegraphics[width=\linewidth,height=0.53\linewidth]{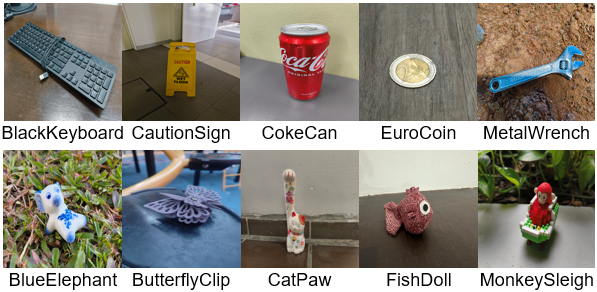}
  \caption{One example image for each of the 10 concepts contained within the NovelConcepts10 dataset.}
  \label{fig:Novel_Concepts}
\end{figure}

The NovelConcepts10 dataset consists of five images for each of 10 distinct concepts, for a total of 50 images. Each object is located roughly at the center of each image, as is common in most LAION \cite{2022_Schuhmann_LAION} and ImageNet \cite{2009_Deng_ImageNet} images. Angle (or pose) and background are different for each image of a single object. The images are resized to 512x512 and stored in PNG format to preserve quality. When choosing concepts to capture, we included both non-unique objects (e.g., CokeCan) and unique objects (e.g., FishDoll). Non-unique objects are easily recognizable and are likely included in large training datasets whereas as unique objects are rare and unlikely to be captured in most training datasets. Fig. \ref{fig:Novel_Concepts} shows an example of each concept in the NovelConcepts10 dataset.

\section{Poison Descriptions}
\label{sec:appendix_poisons}

\begin{table*}[t]
  \centering
  \caption{NovelConcepts10 DINOv2 Similarity for various TI hyperparameter settings. Bold values indicate best settings per poison type.}
  \begin{tabular}{r|c|ccc|ccc|ccc}
    \toprule
    \multicolumn{2}{r|}{LR Sched.} & \multicolumn{3}{|c|}{Constant} & \multicolumn{3}{|c|}{Linear} & \multicolumn{3}{|c}{Cosine} \\
    \multicolumn{2}{r|}{LR} & \multicolumn{1}{|c|}{$5e^{-5}$} & \multicolumn{1}{|c|}{$5e^{-4}$} & \multicolumn{1}{|c|}{$5e^{-3}$} & \multicolumn{1}{|c|}{$5e^{-5}$} & \multicolumn{1}{|c|}{$5e^{-4}$} & \multicolumn{1}{|c|}{$5e^{-3}$} & \multicolumn{1}{|c|}{$5e^{-5}$} & \multicolumn{1}{|c|}{$5e^{-4}$} & $5e^{-3}$ \\
    \midrule
    \multirow{3}{*}{\begin{turn}{90}Clean\end{turn}} 
    & 1k & 0.14 & 0.36 & 0.43 & 0.11 & 0.31 & 0.41 & 0.11 & 0.36 & 0.41 \\
    & 5k & 0.33 & 0.44 & 0.47 & 0.28 & 0.42 & 0.47 & 0.26 & 0.40 & 0.48 \\
    & 10k & 0.37 & 0.44 & 0.49 & 0.33 & 0.44 & 0.50 & 0.36 & 0.43 & \textbf{0.51} \\
    \midrule
    \multirow{3}{*}{\begin{turn}{90}ADM+\end{turn}} 
    & 1k & 0.11 & 0.09 & 0.08 & 0.11 & 0.10 & 0.10 & 0.11 & \textbf{0.12} & 0.11 \\
    & 5k & 0.10 & 0.10 & 0.08 & 0.10 & 0.09 & 0.09 & 0.10 & 0.09 & 0.08 \\
    & 10k & 0.10 & 0.10 & 0.09 & 0.11 & 0.11 & 0.09 & 0.11 & 0.11 & 0.09 \\
    \midrule
    \multirow{3}{*}{\begin{turn}{90}ADM-\end{turn}} 
    & 1k & 0.11 & 0.29 & 0.30 & 0.11 & 0.24 & 0.29 & 0.11 & 0.25 & 0.28 \\
    & 5k & 0.22 & 0.29 & 0.33 & 0.21 & 0.30 & 0.30 & 0.20 & 0.29 & 0.34 \\
    & 10k & 0.22 & 0.26 & 0.37 & 0.24 & 0.29 & 0.35 & 0.24 & 0.28 & \textbf{0.40} \\
    \midrule
    \multirow{3}{*}{\begin{turn}{90}EA\end{turn}} 
    & 1k & \textbf{0.11} & 0.09 & 0.07 & \textbf{0.11} & \textbf{0.11} & 0.07 & \textbf{0.11} & 0.10 & 0.06 \\
    & 5k & \textbf{0.11} & 0.08 & 0.08 & 0.10 & 0.09 & 0.07 & 0.10 & 0.07 & 0.07 \\
    & 10k & 0.09 & 0.09 & 0.07 & 0.10 & 0.08 & 0.09 & 0.10 & 0.09 & 0.09 \\
    \bottomrule
  \end{tabular}
  \label{tab:Hyperparams_DINOv2}
\end{table*}

We analyze multiple adversarial methods. In particular, we use AdvDM (ADM+) \cite{2023_Liang_AdvDM}, SDS (SDS+) \cite{2024_Xue_SDS}, as well as EncoderAttack (EA) and DiffusionAttack (DA) \cite{2023_Salman_Photoguard}. We implement ADM+ and EA via attack modes 0 and 1 from the MIST library \cite{2023_Liang_Mist}. We also analyze the gradient descent versions of ADM+ and SDS+, ADM- and SDS- respectively, since gradient descent poisons are also effective \cite{2024_Xue_SDS}. Table \ref{tab:Poison_Descriptions} describes the attack direction, targeting type, and objective of each poison. We restrict each poison perturbation to a $l^{\infty}$ bound of size $\kappa=16/256$. For ADM+, ADM-, SDS+, SDS-, and EA, we apply $100$ projected gradient steps of strength $\eta=1/256$. For DA, we follow the default sampling and optimization settings. We use the high-contrast MIST image from the MIST library as the target image for EA, which was cited by \cite{2023_Liang_Mist} as a better target than gray images.

\begin{table}[t]
  \centering
  \caption{Characteristics of examined poisons.}
  \begin{tabular}{lccc}
    \toprule
    Poison & Direction & Targeted & Objective \\
    \midrule
    ADM+ \cite{2023_Liang_AdvDM} & $\uparrow$ & No & DM Loss \\
    ADM- & $\downarrow$ & No & DM Loss \\
    SDS+ \cite{2024_Xue_SDS} & $\uparrow$ & No & DM Loss \\
    SDS- \cite{2024_Xue_SDS} & $\downarrow$ & No & DM Loss \\
    EA \cite{2023_Salman_Photoguard} & $\downarrow$ & Yes & Encoding Dist \\
    DA \cite{2023_Salman_Photoguard} & $\uparrow$ & No & DM Sampling \\
    \bottomrule
  \end{tabular}
  \label{tab:Poison_Descriptions}
\end{table}

\section{Hyperparameters Ablation}
\label{sec:appendix_hyperparams}

We ablate the hyperparameter settings (learning rate, learning rate schedule, training steps) for TI training on clean and poisoned NovelConcepts10 datasets. Results are shown in Tables \ref{tab:Hyperparams_DINOv2}, \ref{tab:Hyperparams_FID}, and \ref{tab:Hyperparams_CLIPScore} (we apologize for the nonconsecutive table order - the formatting for this section was difficult).

In general, we find that high learning rates and more training steps benefit performance for Clean and ADM- datasets, but reduce performance on ADM+ and EA. Low learning rates and training steps can inhibit learning (of both clean and adversarial signals). The default learning rate of $5e^{-4}$ yields balanced performance across all datasets and thus we focus on this setting. At a learning rate of $5e^{-4}$, a constant learning rate schedule generally gives best performance. Finally, $5000$ training steps outperforms $1000$ training steps across most poisoned datasets, and the marginal improvements (if any) seen with $10000$ training steps are not worth the extended training time.

We note additional interesting behaviors from the hyperparameter ablation. In particular, we find that training for $1000$ is almost equally affected by poisoning as training for $10000$ steps, even as clean performance increases. This suggests that early stopping cannot avoid adversarial signals. Likewise, a higher learning rate boosts performance for clean images but reduces performance on concepts poisoned by ADM+ or EA. Using linear or cosine decay reduces performance when training for $1000$ steps but does not significantly impact results for $5000$ or $10000$ steps. Finally, it appears that ADM- is the weakest poison across a wide range of settings.

\clearpage
\begin{table*}[h]
  \centering
  \caption{NovelConcepts10 FID for various TI hyperparameter settings. Best settings per poison in bold.}
  \begin{tabular}{r|c|ccc|ccc|ccc}
    \toprule
    \multicolumn{2}{r|}{LR Sched.} & \multicolumn{3}{|c|}{Constant} & \multicolumn{3}{|c|}{Linear} & \multicolumn{3}{|c}{Cosine} \\
    \multicolumn{2}{r|}{LR} & \multicolumn{1}{|c|}{$5e^{-5}$} & \multicolumn{1}{|c|}{$5e^{-4}$} & \multicolumn{1}{|c|}{$5e^{-3}$} & \multicolumn{1}{|c|}{$5e^{-5}$} & \multicolumn{1}{|c|}{$5e^{-4}$} & \multicolumn{1}{|c|}{$5e^{-3}$} & \multicolumn{1}{|c|}{$5e^{-5}$} & \multicolumn{1}{|c|}{$5e^{-4}$} & $5e^{-3}$ \\
    \midrule
    \multirow{3}{*}{\begin{turn}{90}Clean\end{turn}} 
    & 1k & 397 & 307 & 285 & 415 & 330 & 289 & 414 & 301 & 296 \\
    & 5k & 317 & 285 & 266 & 335 & 287 & 273 & 341 & 297 & 270 \\
    & 10k & 301 & 285 & 269 & 315 & 285 & 262 & 305 & 286 & \textbf{258} \\
    \midrule
    \multirow{3}{*}{\begin{turn}{90}ADM+\end{turn}} 
    & 1k & 417 & 449 & 457 & 412 & 446 & 446 & \textbf{411} & 434 & 447 \\
    & 5k & 440 & 453 & 459 & 439 & 452 & 453 & 436 & 447 & 459 \\
    & 10k & 431 & 449 & 457 & 433 & 438 & 454 & 438 & 447 & 454 \\
    \midrule
    \multirow{3}{*}{\begin{turn}{90}ADM-\end{turn}} 
    & 1k & 414 & 344 & 337 & 416 & 359 & 341 & 414 & 348 & 343 \\
    & 5k & 370 & 339 & 333 & 368 & 341 & 337 & 373 & 348 & 323 \\
    & 10k & 370 & 358 & 318 & 358 & 347 & 324 & 365 & 352 & \textbf{300} \\
    \midrule
    \multirow{3}{*}{\begin{turn}{90}EA\end{turn}} 
    & 1k & \textbf{414} & 435 & 440 & \textbf{414} & 418 & 444 & \textbf{414} & 424 & 448 \\
    & 5k & 419 & 443 & 457 & 416 & 437 & 450 & 416 & 446 & 467 \\
    & 10k & 425 & 440 & 465 & 424 & 442 & 448 & 417 & 439 & 457 \\
    \bottomrule
  \end{tabular}
  \label{tab:Hyperparams_FID}
\end{table*}

\begin{table*}[h]
  \centering
  \caption{NovelConcepts10 CLIP Score for various TI hyperparameter settings. Best settings per poison in bold.}
  \begin{tabular}{r|c|ccc|ccc|ccc}
    \toprule
    \multicolumn{2}{r|}{LR Sched.} & \multicolumn{3}{|c|}{Constant} & \multicolumn{3}{|c|}{Linear} & \multicolumn{3}{|c}{Cosine} \\
    \multicolumn{2}{r|}{LR} & \multicolumn{1}{|c|}{$5e^{-5}$} & \multicolumn{1}{|c|}{$5e^{-4}$} & \multicolumn{1}{|c|}{$5e^{-3}$} & \multicolumn{1}{|c|}{$5e^{-5}$} & \multicolumn{1}{|c|}{$5e^{-4}$} & \multicolumn{1}{|c|}{$5e^{-3}$} & \multicolumn{1}{|c|}{$5e^{-5}$} & \multicolumn{1}{|c|}{$5e^{-4}$} & $5e^{-3}$ \\
    \midrule
    \multirow{3}{*}{\begin{turn}{90}Clean\end{turn}} 
    & 1k & 0.37 & 0.48 & 0.50 & 0.34 & 0.47 & 0.46 & 0.34 & 0.50 & \textbf{0.51} \\
    & 5k & 0.47 & \textbf{0.51} & 0.46 & 0.45 & \textbf{0.51} & 0.48 & 0.44 & \textbf{0.51} & \textbf{0.51} \\
    & 10k & 0.50 & 0.49 & 0.43 & 0.46 & \textbf{0.51} & 0.48 & 0.50 & 0.47 & 0.48 \\
    \midrule
    \multirow{3}{*}{\begin{turn}{90}ADM+\end{turn}} 
    & 1k & 0.40 & 0.50 & 0.46 & 0.35 & \textbf{0.52} & 0.48 & 0.35 & 0.49 & 0.51 \\
    & 5k & 0.46 & 0.46 & 0.42 & 0.48 & 0.46 & 0.44 & 0.46 & 0.48 & 0.45 \\
    & 10k & 0.45 & 0.45 & 0.40 & 0.48 & 0.46 & 0.44 & 0.49 & 0.45 & 0.44 \\
    \midrule
    \multirow{3}{*}{\begin{turn}{90}ADM-\end{turn}} 
    & 1k & 0.38 & 0.50 & 0.47 & 0.37 & 0.46 & \textbf{0.51} & 0.37 & 0.45 & 0.50 \\
    & 5k & 0.48 & 0.45 & 0.45 & 0.44 & 0.48 & 0.45 & 0.43 & 0.45 & 0.47 \\
    & 10k & 0.45 & 0.46 & 0.41 & 0.46 & 0.43 & 0.45 & 0.46 & 0.47 & 0.43 \\
    \midrule
    \multirow{3}{*}{\begin{turn}{90}EA\end{turn}} 
    & 1k & 0.37 & \textbf{0.49} & 0.46 & 0.35 & 0.44 & 0.46 & 0.36 & 0.45 & 0.46 \\
    & 5k & 0.45 & 0.45 & 0.43 & 0.39 & 0.45 & 0.43 & 0.39 & 0.46 & 0.45 \\
    & 10k & 0.44 & 0.44 & 0.38 & 0.43 & 0.44 & 0.43 & 0.42 & 0.44 & 0.40 \\
    \bottomrule
  \end{tabular}
  \label{tab:Hyperparams_CLIPScore}
\end{table*}

\clearpage
\section{Timestep Range Ablation}
\label{sec:appendix_trange}

\begin{figure}[h]
  \centering
  \includegraphics[width=0.9\linewidth]{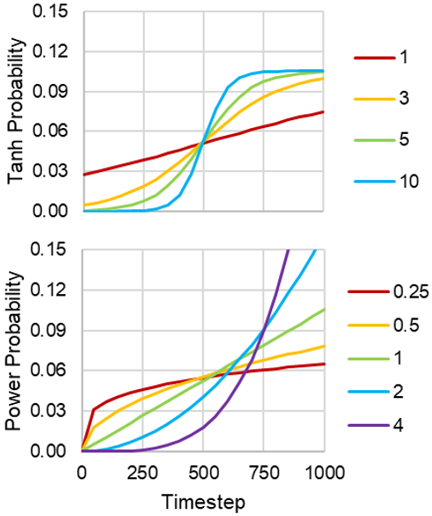}
  \caption{Probability curves for timestep sampling, displaying $tanh$ distributions (top) and power distributions (bottom). Distribution parameters for each curve are given in the legends.}
  \label{fig:Prob_Curves}
\end{figure}

We ablate methods of restricting training to higher timesteps. We investigate high thresholding ($t \sim \mathcal{U}(\rho,1)$), power distributions ($p(t) \propto t^{\rho}$), and tanh distributions ($p(t) \propto tanh(\rho(t-0.5))/2+0.5$). For comparison, we also evaluate low thresholding ($t \sim \mathcal{U}(0,\rho)$). We abuse notation and use $\rho$ for various function parameters that control the shape of each sampling distribution; increasing $\rho$ increases sampling probability for higher timesteps. Here, $t$ is sampled in domain $[0,1]$ and then the sampled output is rescaled to $[0,1000]$ during training. Fig. \ref{fig:Prob_Curves} displays the tanh and power probability distributions for various $\rho$ values.

It can be seen from ablation results in Tables \ref{tab:tRange_DINOv2}, \ref{tab:tRange_FID}, and \ref{tab:tRange_CLIPScore} that performance on datasets poisoned by ADM+ improves significantly as timestep sampling shifts towards higher timesteps. Performance on EA-poisoned concepts also increases slightly while performance on ADM- is relatively stable across methods. Simple high-thresholding is the often the best timestep restriction method. In all cases, we note a performance decrease when t is concentrated at extremely high timesteps (e.g., $t \ge 900$) since learning true features in this high-noise range is challenging. Lastly, we empirically validate the hypothesis that adversarial signals are concentrated at lower-middle timesteps by demonstrating that performance for low thresholding ($t \sim \mathcal{U}(0,\rho)$) is consistently worse than nominal sampling.

\clearpage
\begin{table*}[t]
  \centering
  \caption{NovelConcepts10 DINOv2 Similarity for various timestep range restrictions at various $\rho$ values. Best settings per poison in bold.}
  \begin{tabular}{l|ccccc|ccccc}
    \toprule
    Curve & Nominal & \multicolumn{4}{|c|}{Tanh} & \multicolumn{5}{|c}{Power} \\
    $\rho$ & - & \multicolumn{1}{|c|}{1} & \multicolumn{1}{|c|}{3} & \multicolumn{1}{|c|}{5} & \multicolumn{1}{|c|}{10} & \multicolumn{1}{|c|}{0.25} & \multicolumn{1}{|c|}{0.5} & \multicolumn{1}{|c|}{1} & \multicolumn{1}{|c|}{2} & 4 \\
    \midrule
    Clean & \multicolumn{1}{|c|}{0.44} & 0.44 & 0.47 & \textbf{0.49} & 0.45 & 0.45 & 0.46 & 0.45 & 0.47 & 0.47 \\
    ADM+ & \multicolumn{1}{|c|}{0.10} & 0.13 & 0.11 & 0.20 & 0.18 & 0.10 & 0.10 & 0.12 & 0.24 & 0.27 \\
    ADM- & \multicolumn{1}{|c|}{0.29} & \textbf{0.31} & 0.30 & 0.28 & 0.27 & 0.28 & 0.29 & 0.30 & 0.26 & 0.25 \\
    EA & \multicolumn{1}{|c|}{0.08} & 0.10 & 0.09 & 0.10 & 0.11 & 0.08 & 0.08 & 0.10 & 0.10 & 0.11 \\
    \toprule
    Curve & \multicolumn{5}{|c|}{Threshold Low} & \multicolumn{5}{|c}{Threshold High} \\
    $\rho$ & \multicolumn{1}{|c|}{0.1} & \multicolumn{1}{|c|}{0.2} & \multicolumn{1}{|c|}{0.3} & \multicolumn{1}{|c|}{0.4} & \multicolumn{1}{|c|}{0.5} & \multicolumn{1}{|c|}{0.5} & \multicolumn{1}{|c|}{0.6} & \multicolumn{1}{|c|}{0.7} & \multicolumn{1}{|c|}{0.8} & 0.9 \\
    \midrule
    Clean & 0.17 & 0.24 & 0.29 & 0.31 & 0.37 & 0.46 & 0.44 & 0.39 & 0.37 & 0.31 \\
    ADM+ & 0.08 & 0.08 & 0.08 & 0.10 & 0.09 & 0.22 & 0.21 & 0.28 & \textbf{0.33} & 0.26 \\
    ADM- & 0.13 & 0.16 & 0.20 & 0.27 & 0.26 & 0.26 & 0.25 & 0.23 & 0.19 & 0.18 \\
    EA & 0.07 & 0.07 & 0.08 & 0.07 & 0.07 & 0.11 & 0.13 & \textbf{0.14} & 0.13 & 0.10 \\
    \bottomrule
  \end{tabular}
  \label{tab:tRange_DINOv2}
\end{table*}

\begin{table*}[h]
  \centering
  \caption{NovelConcepts10 FID for various timestep range restrictions at various $\rho$ values. Best settings per poison in bold.}
  \begin{tabular}{l|ccccc|ccccc}
    \toprule
    Curve & Nominal & \multicolumn{4}{|c|}{Tanh} & \multicolumn{5}{|c}{Power} \\
    $\rho$ & - & \multicolumn{1}{|c|}{1} & \multicolumn{1}{|c|}{3} & \multicolumn{1}{|c|}{5} & \multicolumn{1}{|c|}{10} & \multicolumn{1}{|c|}{0.25} & \multicolumn{1}{|c|}{0.5} & \multicolumn{1}{|c|}{1} & \multicolumn{1}{|c|}{2} & 4 \\
    \midrule
    Clean & \multicolumn{1}{|c|}{285} & 288 & 271 & \textbf{265} & 283 & 281 & 274 & 279 & 270 & \textbf{265} \\
    ADM+ & \multicolumn{1}{|c|}{453} & 440 & 443 & 401 & 399 & 448 & 448 & 430 & 376 & 360 \\
    ADM- & \multicolumn{1}{|c|}{339} & \textbf{335} & 341 & 344 & 351 & 351 & 344 & \textbf{335} & 349 & 362 \\
    EA & \multicolumn{1}{|c|}{443} & 445 & 440 & 439 & 432 & 442 & 442 & 433 & 425 & 424 \\
    \toprule
    Curve & \multicolumn{5}{|c|}{Threshold Low} & \multicolumn{5}{|c}{Threshold High} \\
    $\rho$ & \multicolumn{1}{|c|}{0.1} & \multicolumn{1}{|c|}{0.2} & \multicolumn{1}{|c|}{0.3} & \multicolumn{1}{|c|}{0.4} & \multicolumn{1}{|c|}{0.5} & \multicolumn{1}{|c|}{0.5} & \multicolumn{1}{|c|}{0.6} & \multicolumn{1}{|c|}{0.7} & \multicolumn{1}{|c|}{0.8} & 0.9 \\
    \midrule
    Clean & 387 & 365 & 342 & 334 & 323 & 279 & 285 & 304 & 312 & 336 \\
    ADM+ & 442 & 448 & 449 & 451 & 447 & 388 & 389 & 352 & \textbf{336} & 364 \\
    ADM- & 426 & 407 & 387 & 360 & 363 & 351 & 361 & 369 & 393 & 400 \\
    EA & 437 & 441 & 445 & 443 & 451 & 424 & 414 & \textbf{413} & \textbf{413} & 439 \\
    \bottomrule
  \end{tabular}
  \label{tab:tRange_FID}
\end{table*}

\begin{table*}[h]
  \centering
  \caption{NovelConcepts10 CLIP Score for various timestep range restrictions at various $\rho$ values. Best settings per poison in bold.}
  \begin{tabular}{l|ccccc|ccccc}
    \toprule
    Curve & Nominal & \multicolumn{4}{|c|}{Tanh} & \multicolumn{5}{|c}{Power} \\
    $\rho$ & - & \multicolumn{1}{|c|}{1} & \multicolumn{1}{|c|}{3} & \multicolumn{1}{|c|}{5} & \multicolumn{1}{|c|}{10} & \multicolumn{1}{|c|}{0.25} & \multicolumn{1}{|c|}{0.5} & \multicolumn{1}{|c|}{1} & \multicolumn{1}{|c|}{2} & 4 \\
    \midrule
    Clean & \multicolumn{1}{|c|}{0.51} & \textbf{0.53} & 0.51 & 0.50 & 0.52 & 0.51 & 0.51 & \textbf{0.53} & 0.47 & 0.50 \\
    ADM+ & \multicolumn{1}{|c|}{0.46} & 0.49 & 0.46 & \textbf{0.53} & 0.50 & 0.47 & 0.47 & 0.47 & 0.49 & 0.52 \\
    ADM- & \multicolumn{1}{|c|}{0.45} & 0.47 & \textbf{0.49} & 0.47 & 0.44 & \textbf{0.49} & 0.48 & \textbf{0.49} & 0.47 & 0.44 \\
    EA & \multicolumn{1}{|c|}{0.45} & 0.47 & 0.46 & 0.45 & 0.44 & 0.47 & 0.46 & 0.46 & 0.48 & 0.45 \\
    \toprule
    Curve & \multicolumn{5}{|c|}{Threshold Low} & \multicolumn{5}{|c}{Threshold High} \\
    $\rho$ & \multicolumn{1}{|c|}{0.1} & \multicolumn{1}{|c|}{0.2} & \multicolumn{1}{|c|}{0.3} & \multicolumn{1}{|c|}{0.4} & \multicolumn{1}{|c|}{0.5} & \multicolumn{1}{|c|}{0.5} & \multicolumn{1}{|c|}{0.6} & \multicolumn{1}{|c|}{0.7} & \multicolumn{1}{|c|}{0.8} & 0.9 \\
    \midrule
    Clean & 0.44 & 0.45 & 0.43 & 0.49 & 0.49 & 0.51 & 0.49 & 0.50 & 0.49 & 0.45 \\
    ADM+ & 0.38 & 0.41 & 0.43 & 0.44 & 0.44 & 0.49 & 0.50 & 0.48 & 0.48 & 0.47 \\
    ADM- & 0.44 & 0.45 & 0.45 & 0.48 & 0.46 & 0.44 & 0.45 & 0.48 & \textbf{0.49} & 0.48 \\
    EA & 0.41 & 0.43 & 0.40 & 0.43 & 0.45 & 0.44 & 0.46 & 0.48 & \textbf{0.49} & \textbf{0.49} \\
    \bottomrule
  \end{tabular}
  \label{tab:tRange_CLIPScore}
\end{table*}

\clearpage
\section{Masking Ablation}
\label{sec:appendix_formulation_Masking}

\begin{table}[h]
  \centering
  \caption{NovelConcepts10 DINOv2 Similarity for different masking methods.}
  \begin{tabular}{lccccc}
    \toprule
    Poison & Nominal & LM & IM & LIM & ZM \\
    \midrule
    Clean & 0.44 & \textbf{0.45} & 0.38 & 0.42 & 0.20 \\
    \midrule
    ADM+ & 0.10 & \textbf{0.36} & 0.28 & 0.33 & 0.11 \\
    ADM- & 0.29 & \textbf{0.41} & 0.31 & 0.31 & 0.22 \\
    SDS+ & 0.11 & \textbf{0.33} & 0.30 & 0.26 & 0.14 \\
    SDS- & 0.25 & \textbf{0.39} & 0.29 & 0.29 & 0.16 \\
    EA & 0.08 & \textbf{0.30} & 0.29 & 0.26 & 0.15 \\
    DA & 0.27 & \textbf{0.38} & 0.34 & 0.31 & 0.16 \\
    \midrule
    Psn Avg & 0.18 & \textbf{0.36} & 0.30 & 0.29 & 0.16 \\
    \bottomrule
  \end{tabular}
  \label{tab:masking_ablation_DINO}
\end{table}

\begin{table}[h]
  \centering
  \caption{NovelConcepts10 FID for different masking methods.}
  \begin{tabular}{lccccc}
    \toprule
    Poison & Nominal & LM & IM & LIM & ZM \\
    \midrule
    Clean & \textbf{285} & \textbf{285} & 299 & 293 & 406 \\
    \midrule
    ADM+ & 453 & \textbf{309} & 344 & 342 & 439 \\
    ADM- & 339 & \textbf{299} & 330 & 341 & 378 \\
    SDS+ & 439 & \textbf{341} & \textbf{341} & 365 & 428 \\
    SDS- & 352 & \textbf{304} & 346 & 342 & 405 \\
    EA & 443 & 359 & \textbf{348} & 365 & 421 \\
    DA & 352 & \textbf{309} & 328 & 333 & 434 \\
    \midrule
    Psn Avg & 396 & \textbf{320} & 340 & 348 & 417 \\
    \bottomrule
  \end{tabular}
  \label{tab:masking_ablation_FID}
\end{table}

\begin{table}[h]
  \centering
  \caption{NovelConcepts10 CLIP Score for different masking methods.}
  \begin{tabular}{lccccc}
    \toprule
    Poison & Nominal & LM & IM & LIM & ZM \\
    \midrule
    Clean & 0.51 & \textbf{0.52} & 0.48 & \textbf{0.52} & 0.46 \\
    \midrule
    ADM+ & 0.46 & 0.47 & 0.48 & \textbf{0.52} & 0.42 \\
    ADM- & 0.45 & 0.54 & 0.50 & \textbf{0.51} & 0.44 \\
    SDS+ & 0.44 & \textbf{0.48} & \textbf{0.48} & \textbf{0.48} & 0.45 \\
    SDS- & 0.45 & 0.49 & 0.49 & \textbf{0.50} & 0.41 \\
    EA & 0.45 & \textbf{0.50} & 0.49 & 0.48 & 0.42 \\
    DA & 0.49 & 0.48 & 0.52 & \textbf{0.53} & 0.46 \\
    \midrule
    Psn Avg & 0.46 & 0.49 & 0.49 & \textbf{0.50} & 0.43 \\
    \bottomrule
  \end{tabular}
  \label{tab:masking_ablation_CLIPScore}
\end{table}

We define the various objectives used by loss masking (LM), input masking (IM), loss-input masking (LIM), and latent masking (ZM). The notation here follows that of Sections \ref{sec:preliminaries} and \ref{subsec:SZT_Method}. The LM objective is given by
\begin{equation}
    \label{eqn:L_LM}
    L_{LM} (x, t, c, M_{x}) = ||(\epsilon_{\theta}(z_{t},t,c) -\epsilon) \odot M_{z}||_{2}^2,
\end{equation}
which is similar to the SZT objective from Eq. \ref{eqn:SZT}. IM applies masking only to the input $x$ and is given by
\begin{equation}
    \label{eqn:L_IM}
    L_{IM} (x, t, c, M_{x}) = ||\epsilon_{\theta}(z_{tM},t,c) -\epsilon||_{2}^2,
\end{equation}
where $z_{tM}$ is the noised latent of a masked input image, $z_{tM} = \sqrt{\overline{\alpha}_t}\mathcal{E}(x \odot M_x) + \sqrt{1- \overline{\alpha}_t}\epsilon$. LIM combines the $L_{LM}$ and $L_{IM}$ objectives as
\begin{equation}
    L_{LIM} (x, t, c, M_{x}) = ||(\epsilon_{\theta}(z_{tM},t,c)-\epsilon) \odot M_{z}||_{2}^2.
\end{equation}
ZM applies masking to the latent vector $z_t$ and is given by
\begin{equation}
    L_{ZM} (x, t, c, M_{x}) = ||\epsilon_{\theta}(z_{t} \odot M_{z},t,c) -\epsilon||_{2}^2.
\end{equation}

We evaluate masking types on NovelConcepts10 across all poisons. Tables \ref{tab:masking_ablation_DINO}, \ref{tab:masking_ablation_FID}, and \ref{tab:masking_ablation_CLIPScore} demonstrate that LM outperforms all other forms of masking for poison defense, as it is the only method that fully preserves background information in the forward process. IM performs underwhelmingly and LIM is apparently limited by the input mask applied by IM. ZM consistently gives the lowest performance.

\begin{figure}[b]
  \centering
  \includegraphics[width=\linewidth]{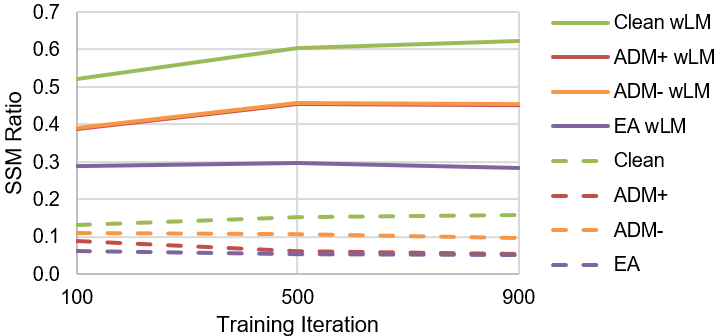}
  \caption{Average $R_{n, M_{z}}$ for clean and poisoned versions of NovelConcepts10 throughout TI training. ``wLM" denotes TI with loss masking.}
  \label{fig:SSM_Ratio_Masking}
\end{figure}

To further evaluate the impact of LM, we measure the proportion of SSM values within novel concept regions compared to the sum of all SSM values throughout the image. This metric can be captured using the ratio defined below, with notation mirroring that of Sec. \ref{subsec:Semantic_Sensitivity_Map}:
\begin{equation}
    \label{eqn:SSM_Ratio}
    R_{n, M_{z}} = \frac{\sum_{i,j}(SSM(x, t, \widehat{e}, n) \odot M_{z})_{i,j}}{\sum_{i,j}SSM(x, t, \widehat{e}, n)_{i,j}}.
\end{equation}

We measure $R_{n, M_z}$ for clean and poisoned versions of NovelConcepts10 at steps $100$, $500$, and $900$ during TI training and average the values across all concepts. The results in Fig. \ref{fig:SSM_Ratio_Masking} show that the ratio of SSM within novel concept regions naturally increases throughout training for clean data. For poisoned datasets without LM, the proportion of SSM in the novel concept regions never increases, indicating that learning is distracted away from the novel concept regions by adversarial signals. Only with LM does TI focus on the novel concept regions for poisoned datasets.

\clearpage
\section{Mask Dilation Ablation}
\label{sec:appendix_maskdilate}

In initial experiments with SZT, we found that combining T600 and LM tended to perform below expectations. Although the performance of T600+LM was typically better than established defenses like Regen and AdvClean, it underperformed other SZT ablations like JPEG+T600 or JPEG+LM. We hypothesize that strong restrictions in both time (i.e., T600) and space (i.e., LM) may be too restrictive and may hinder concept learning during TI training. Therefore, we evaluated additional configurations of SZT that ease temporal and spatial restrictions when used in combination. In particular, we investigate lower timestep threshold (e.g., $t \ge 500$, denoted as ``T500") combined with dilated concept masks. To implement mask dilations for LM, we apply the \texttt{ImageFilter.MaxFilter} method from PIL with $8$, $16$, or $24$ pixels of dilation to the $512$x$512$ binary masks, then rescale to $64$x$64$ for latent space. Intuitively, applying mask dilation includes extra background information outside of the novel concept region during loss backpropagation.

Tables \ref{tab:NC10_Dilate_DINOv2}, \ref{tab:NC10_Dilate_FID}, and \ref{tab:NC10_Dilate_CLIPScore} display the DINOv2 similarity, FID, and CLIP Score for various combinations of T500, T600, and dilated masks (denoted ``LM-D08", ``LM-D16", and ``LM-D24"). Of the various LM configurations, $16$ pixels of dilation (LM-D16) demonstrates the highest robustness to poisons. Furthermore, when combining T500 or T600 with LM, using T500+LM-D16 gives the best performance, supporting our ``too restrictive" hypothesis above. Our final implementation of SZT uses JPEG preprocessing with T500+LM-D16 and further improves poison defense beyond all other ablations.

\section{Defense Comparison for CustomConcept101}
\label{sec:appendix_CC101_baselines}

We additionally include FID and CLIP Score metrics for CustomConcept101 in Tables \ref{tab:CC101_Baseline_FID} and \ref{tab:CC101_Baseline_CLIPScore}, complementing the DINOv2 Similarity results in Table \ref{tab:CC101_Baseline_DINOv2}. Trends in defense methods are generally similar as those observed in Sec. \ref{subsubsec:SZT_Performance} As observed for the DINOv2 Similarity results, SZT is the best method for poison defense.

\begin{figure}[t]
  \centering
  \includegraphics[width=\linewidth]{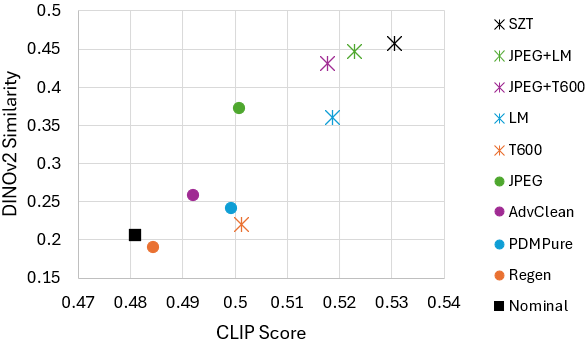}
  \caption{DINOv2 Similarity versus CLIP Score for all defenses on CustomConcept101. Values are from the ``Psn Avg" rows in Tables \ref{tab:CC101_Baseline_DINOv2} and \ref{tab:CC101_Baseline_CLIPScore}.}
  \label{fig:DINO_CLIP}
\end{figure}

For emphasis, we plot the DINOv2 Similarity versus CLIP Score values for ``Psn Avg" on CustomConcept101 for all defenses in Fig. \ref{fig:DINO_CLIP}. As discussed in Sec. \ref{subsubsec:SZT_Performance}, Regen, PDMPure, and AdvClean all offer minor improvements in poison performance. We note the that despite a drastic improvement in DINOv2 Similarity, JPEG is limited in its prompt fidelity (measured by CLIP Score). This aligns with qualitative observations of its limited concept learning from Fig. \ref{fig:GenImgs_Main}. SZT improves DINOv2 Similarity and CLIP Score beyond all existing defenses, and most ablations of SZT also perform well. We note that all ablations of SZT can beat existing defenses in at least one metric.

\section{Defense Comparison for NovelConcepts10}
\label{sec:appendix_NC10_baselines}

We repeat the same study as Sections \ref{subsubsec:SZT_Performance} and \ref{sec:appendix_CC101_baselines} for NovelConcepts10 and report results in Tables \ref{tab:NC10_Baseline_DINOv2}, \ref{tab:NC10_Baseline_FID}, and \ref{tab:NC10_Baseline_CLIPScore}. All trends for NovelConcepts10 generally mirror those seen for CustomConcept101, validating our prior observations.

\section{Generated Images after TI}
\label{sec:appendix_gen_imgs}

\begin{figure*}[th]
  \centering
  \includegraphics[width=0.97\linewidth]{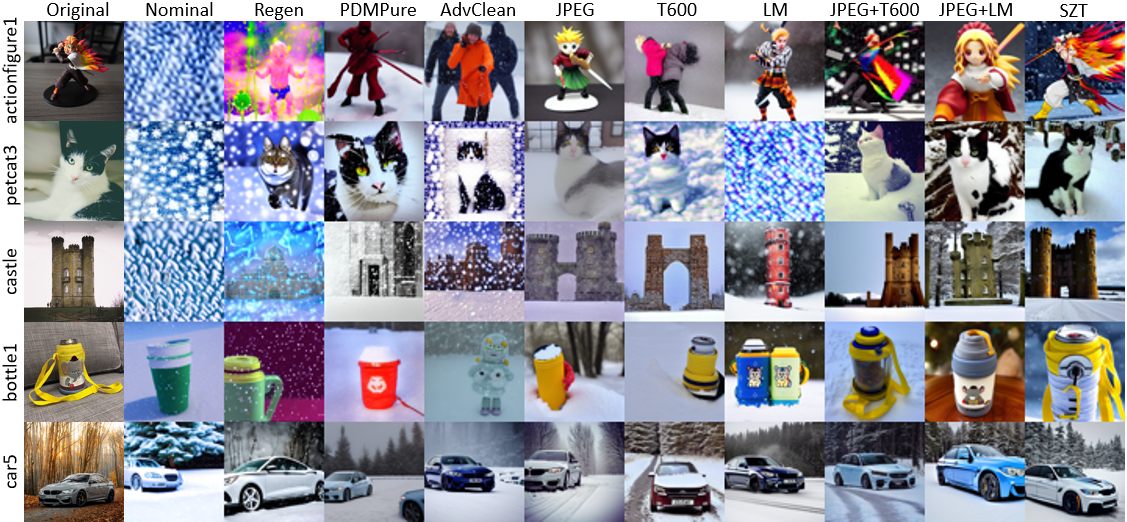}
  \caption{Images generated after TI on various concepts from CustomConcept101 poisoned by ADM+. Prompt: ``a $R*$ in the snow"}
  \label{fig:GenImgs_CC101_ADM}
\end{figure*}

\begin{figure*}[th]
  \centering
  \includegraphics[width=0.97\linewidth]{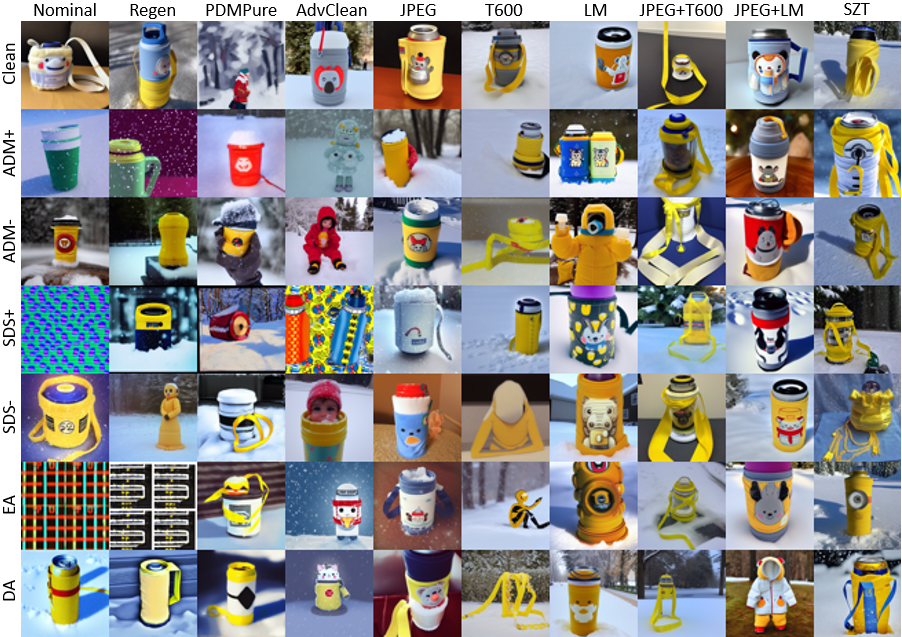}
  \caption{Images generated after TI on things-bottle1 (CustomConcept101) for various poisons. Prompt: ``a $R*$ in the snow"}
  \label{fig:GenImgs_CC101_Bottle1}
\end{figure*}

\begin{figure*}[th]
  \centering
  \includegraphics[width=0.97\linewidth]{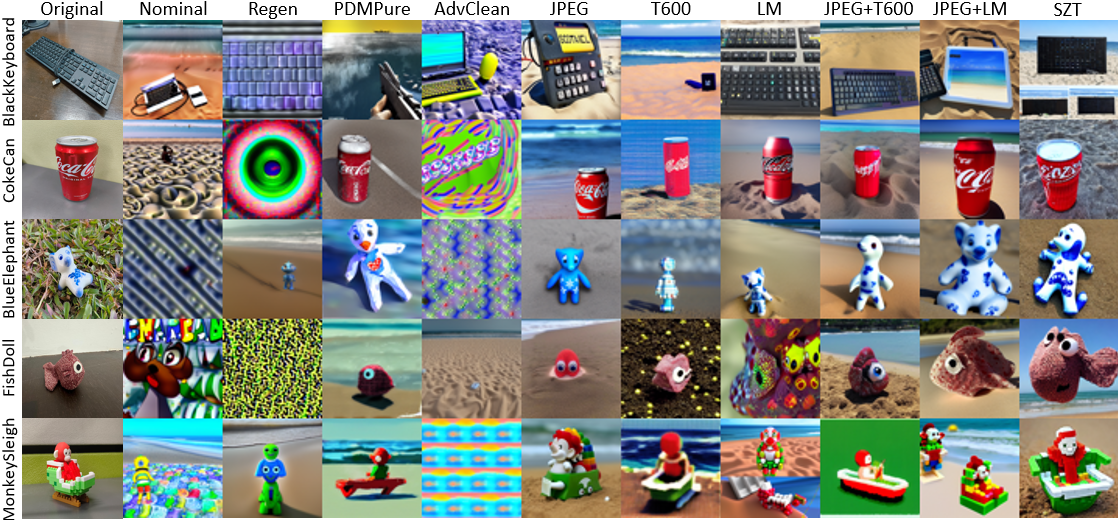}
  \caption{Images generated after TI on various concepts from NovelConcepts10 poisoned by ADM+. Prompt: ``a $R*$ on the beach"}
  \label{fig:GenImgs_NC10_ADM}
\end{figure*}

\begin{figure*}[th]
  \centering
  \includegraphics[width=0.97\linewidth]{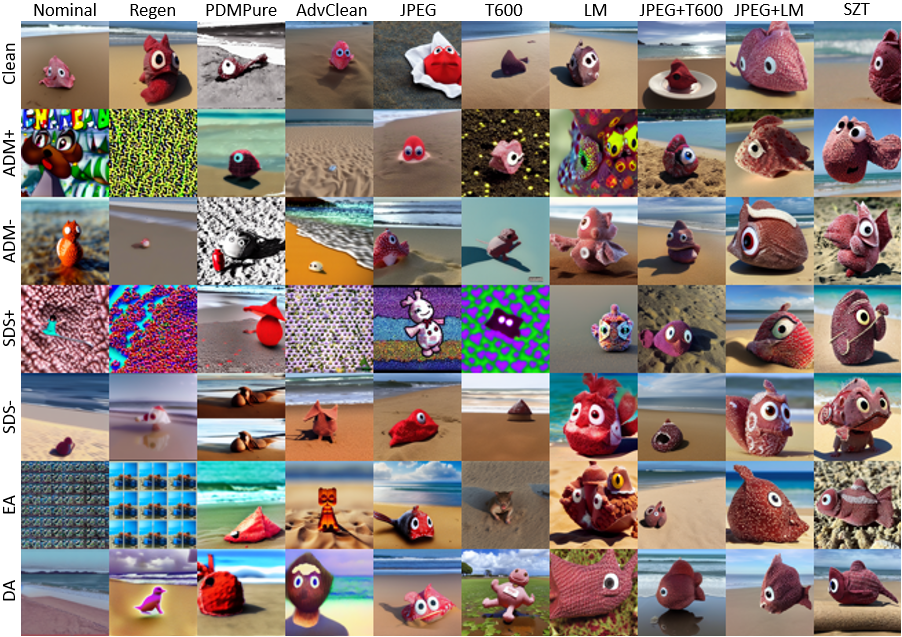}
  \caption{Images generated after TI on FishDoll (NovelConcepts10) for various poisons. Prompt: ``a $R*$ on the beach"}
  \label{fig:GenImgs_NC10_FishDoll}
\end{figure*}

We display generated images after TI for various concepts, poisons, and defenses in figures \ref{fig:GenImgs_CC101_ADM}, \ref{fig:GenImgs_CC101_Bottle1}, \ref{fig:GenImgs_NC10_ADM}, and \ref{fig:GenImgs_NC10_FishDoll}. For each concept/poison/defense, we utilized DINOv2 similarity between generated images and training images to identify the most faithful image for display.  Fig. \ref{fig:GenImgs_CC101_ADM} compares five concepts from CustomConcept101, each poisoned by ADM+. Fig. \ref{fig:GenImgs_CC101_Bottle1} focuses on the things-bottle1 concept from CustomConcept101 across multiple poisons. Figures \ref{fig:GenImgs_NC10_ADM} and \ref{fig:GenImgs_NC10_FishDoll} are analogous, but for NovelConcepts10. For convenience, we also include samples of the original training images in Figures \ref{fig:GenImgs_CC101_ADM} and \ref{fig:GenImgs_NC10_ADM} (left-most column). Observations for all figures here reflect those made for Fig. \ref{fig:GenImgs_Main} in Sec. \ref{subsubsec:SZT_Performance}. 

\begin{table*}
  \centering
  \caption{NovelConcepts10 DINOv2 Similarity for various timestep-restriction and masking ablations.}
  \begin{tabular}{l|c|cccc|cc|cc|cc}
    \toprule
    Defense $\rightarrow$ & Nominal & LM & LM & LM & LM & JPEG & JPEG+ & T500 & T500+ & T600 & T600+ \\
    Poison $\downarrow$ & & & D08 & D16 & D24 & +LM & LM-D16 & +LM & LM-D16 & +LM & LM-D16 \\
    \midrule
    Clean & 0.46 & 0.45 & 0.45 & 0.45 & \textbf{0.47} & 0.43 & 0.46 & 0.44 & 0.46 & 0.44 & 0.46 \\
    \midrule
    ADM+\cite{2023_Liang_AdvDM} & 0.10 & 0.36 & 0.39 & \textbf{0.45} & \textbf{0.45} & 0.43 & 0.44 & 0.39 & 0.40 & 0.33 & 0.36 \\
    ADM- & 0.32 & 0.41 & 0.43 & \textbf{0.44} & 0.42 & 0.43 & 0.42 & 0.25 & 0.31 & 0.22 & 0.24 \\
    SDS+\cite{2024_Xue_SDS} & 0.09 & 0.30 & \textbf{0.44} & \textbf{0.44} & 0.41 & 0.43 & 0.42 & 0.24 & 0.25 & 0.21 & 0.24 \\
    SDS-\cite{2024_Xue_SDS} & 0.11 & 0.33 & 0.40 & \textbf{0.47} & 0.44 & 0.42 & 0.44 & 0.36 & 0.42 & 0.33 & 0.37 \\
    EA\cite{2023_Salman_Photoguard} & 0.25 & 0.39 & 0.41 & 0.41 & \textbf{0.42} & 0.41 & \textbf{0.42} & 0.23 & 0.28 & 0.19 & 0.20 \\
    DA\cite{2023_Salman_Photoguard} & 0.27 & 0.38 & \textbf{0.43} & \textbf{0.43} & 0.42 & 0.42 & 0.42 & 0.29 & 0.34 & 0.33 & 0.34 \\
    \midrule
    Psn Avg & 0.19 & 0.36 & 0.42 & \textbf{0.44} & 0.43 & 0.42 & 0.43 & 0.29 & 0.33 & 0.27 & 0.29 \\
    \bottomrule
  \end{tabular}
  \label{tab:NC10_Dilate_DINOv2}
\end{table*}

\begin{table*}
  \centering
  \caption{NovelConcepts10 FID for various timestep-restriction and masking ablations.}
  \begin{tabular}{l|c|cccc|cc|cc|cc}
    \toprule
    Defense $\rightarrow$ & Nominal & LM & LM & LM & LM & JPEG & JPEG+ & T500 & T500+ & T600 & T600+ \\
    Poison $\downarrow$ & & & D08 & D16 & D24 & +LM & LM-D16 & +LM & LM-D16 & +LM & LM-D16 \\
    \midrule
    Clean & 281 & 285 & 280 & 281 & 281 & 291 & 281 & 284 & 275 & 289 & \textbf{272} \\
    \midrule
    ADM+\cite{2023_Liang_AdvDM} & 451 & 309 & 296 & 279 & \textbf{275} & 291 & 290 & 297 & 284 & 312 & 309 \\
    ADM- & 323 & 299 & 291 & \textbf{287} & 296 & 293 & 302 & 375 & 341 & 374 & 378 \\
    SDS+\cite{2024_Xue_SDS} & 434 & 359 & \textbf{288} & 291 & 302 & 296 & 300 & 372 & 358 & 390 & 381 \\
    SDS-\cite{2024_Xue_SDS} & 439 & 341 & 297 & \textbf{273} & 281 & 299 & 288 & 311 & 283 & 318 & 303 \\
    EA\cite{2023_Salman_Photoguard} & 352 & 304 & 298 & 295 & \textbf{285} & 296 & 297 & 378 & 354 & 381 & 389 \\
    DA\cite{2023_Salman_Photoguard} & 352 & 309 & 295 & \textbf{286} & 295 & 299 & 298 & 350 & 321 & 343 & 328 \\
    \midrule
    Psn Avg & 392 & 320 & 294 & \textbf{285} & 289 & 296 & 296 & 347 & 324 & 353 & 348 \\
    \bottomrule
  \end{tabular}
  \label{tab:NC10_Dilate_FID}
\end{table*}

\begin{table*}
  \centering
  \caption{NovelConcepts10 CLIP Score for various timestep-restriction and masking ablations.}
  \begin{tabular}{l|c|cccc|cc|cc|cc}
    \toprule
    Defense $\rightarrow$ & Nominal & LM & LM & LM & LM & JPEG & JPEG+ & T500 & T500+ & T600 & T600+ \\
    Poison $\downarrow$ & & & D08 & D16 & D24 & +LM & LM-D16 & +LM & LM-D16 & +LM & LM-D16 \\
    \midrule
    Clean & \textbf{0.54} & 0.52 & 0.48 & 0.52 & 0.52 & 0.50 & 0.52 & 0.50 & 0.45 & 0.50 & 0.48 \\
    \midrule
    ADM+\cite{2023_Liang_AdvDM} & 0.46 & 0.47 & 0.48 & 0.50 & 0.48 & 0.51 & 0.52 & 0.52 & \textbf{0.53} & 0.50 & 0.48 \\
    ADM- & 0.49 & \textbf{0.54} & 0.48 & 0.51 & 0.46 & 0.50 & 0.49 & 0.45 & 0.48 & 0.48 & 0.46 \\
    SDS+\cite{2024_Xue_SDS} & 0.45 & 0.50 & 0.50 & 0.48 & 0.48 & 0.47 & 0.47 & 0.49 & 0.51 & \textbf{0.52} & 0.49 \\
    SDS-\cite{2024_Xue_SDS} & 0.44 & 0.48 & 0.46 & 0.51 & 0.50 & 0.50 & \textbf{0.54} & 0.45 & 0.53 & 0.50 & 0.49 \\
    EA\cite{2023_Salman_Photoguard} & 0.45 & 0.49 & 0.47 & 0.46 & 0.46 & 0.51 & \textbf{0.53} & 0.46 & 0.49 & 0.49 & 0.48 \\
    DA\cite{2023_Salman_Photoguard} & 0.49 & 0.48 & 0.47 & 0.46 & 0.48 & \textbf{0.53} & \textbf{0.53} & 0.49 & 0.48 & 0.51 & \textbf{0.53} \\
    \midrule
    Psn Avg & 0.46 & 0.49 & 0.48 & 0.49 & 0.48 & 0.50 & \textbf{0.51} & 0.48 & 0.50 & 0.50 & 0.49 \\
    \bottomrule
  \end{tabular}
  \label{tab:NC10_Dilate_CLIPScore}
\end{table*}

\clearpage

\begin{table*}
  \centering
  \caption{CustomConcept101 FID for various poison defenses.}
  \begin{tabular}{l|c|cccc|ccccc}
    \toprule
    Defense $\rightarrow$ & Nominal & Regen & PDMPure & AdvClean & JPEG & T600 & LM & JPEG & JPEG & SZT \\
    Poison $\downarrow$ & & \cite{2024_Zhao_Regeneration} & \cite{2024_Xue_PDMPure} & \cite{2025_Shidoto_AdverseCleaner} & & & & +T600 & +LM & \\
    \midrule
    Clean & 251 & 255 & 356 & 256 & 257 & \textbf{235} & 250 & 246 & 251 & 245 \\
    \midrule
    ADM+\cite{2023_Liang_AdvDM} & 422 & 386 & 363 & 399 & 292 & 344 & 332 & 269 & 259 & \textbf{257} \\
    ADM- & 302 & 349 & 367 & 297 & 271 & 333 & 267 & 256 & 256 & \textbf{252} \\
    SDS+\cite{2024_Xue_SDS} & 420 & 389 & 357 & 393 & 301 & 348 & 322 & 275 & 261 & \textbf{259} \\
    SDS-\cite{2024_Xue_SDS} & 311 & 357 & 367 & 299 & 273 & 355 & 272 & 259 & 261 & \textbf{248} \\
    EA\cite{2023_Salman_Photoguard} & 413 & 421 & 365 & 367 & 302 & 386 & 329 & 260 & 262 & \textbf{248} \\
    DA\cite{2023_Salman_Photoguard} & 323 & 341 & 347 & 298 & 278 & 358 & 273 & 260 & 255 & \textbf{249} \\
    \midrule
    Psn Avg & 365 & 374 & 361 & 342 & 286 & 354 & 299 & 263 & 259 & \textbf{252} \\
    \bottomrule
  \end{tabular}
  \label{tab:CC101_Baseline_FID}
\end{table*}

\begin{table*}
  \centering
  \caption{CustomConcept101 CLIP Score for various poison defenses.}
  \begin{tabular}{l|c|cccc|ccccc}
    \toprule
    Defense $\rightarrow$ & Nominal & Regen & PDMPure & AdvClean & JPEG & T600 & LM & JPEG & JPEG & SZT \\
    Poison $\downarrow$ & & \cite{2024_Zhao_Regeneration} & \cite{2024_Xue_PDMPure} & \cite{2025_Shidoto_AdverseCleaner} & & & & +T600 & +LM & \\
    \midrule
    Clean & \textbf{0.53} & \textbf{0.53} & 0.48 & 0.52 & 0.52 & \textbf{0.53} & \textbf{0.53} & 0.51 & \textbf{0.53} & \textbf{0.53} \\
    \midrule
    ADM+\cite{2023_Liang_AdvDM} & 0.47 & 0.47 & 0.50 & 0.47 & 0.49 & 0.53 & 0.50 & 0.52 & 0.51 & \textbf{0.54} \\
    ADM- & 0.50 & 0.50 & 0.48 & 0.51 & 0.50 & 0.47 & \textbf{0.52} & 0.51 & \textbf{0.52} & \textbf{0.52} \\
    SDS+\cite{2024_Xue_SDS} & 0.46 & 0.49 & 0.50 & 0.48 & 0.50 & \textbf{0.53} & 0.51 & \textbf{0.53} & \textbf{0.53} & 0.52 \\
    SDS-\cite{2024_Xue_SDS} & 0.48 & 0.48 & 0.50 & 0.51 & 0.50 & 0.46 & \textbf{0.53} & 0.50 & 0.52 & \textbf{0.53} \\
    EA\cite{2023_Salman_Photoguard} & 0.47 & 0.48 & 0.50 & 0.47 & 0.49 & 0.48 & 0.51 & 0.52 & 0.52 & \textbf{0.53} \\
    DA\cite{2023_Salman_Photoguard} & 0.50 & 0.49 & 0.51 & 0.51 & 0.52 & \textbf{0.54} & \textbf{0.54} & 0.52 & 0.53 & \textbf{0.54} \\
    \midrule
    Psn Avg & 0.48 & 0.48 & 0.50 & 0.49 & 0.50 & 0.50 & 0.52 & 0.52 & 0.52 & \textbf{0.53} \\
    \bottomrule
  \end{tabular}
  \label{tab:CC101_Baseline_CLIPScore}
\end{table*}

\clearpage

\begin{table*}
  \centering
  \caption{NovelConcepts10 DINOv2 Similarity for various poison defenses.}
  \begin{tabular}{l|c|cccc|ccccc}
    \toprule
    Defense $\rightarrow$ & Nominal & Regen & PDMPure & AdvClean & JPEG & T600 & LM & JPEG & JPEG & SZT \\
    Poison $\downarrow$ & & \cite{2024_Zhao_Regeneration} & \cite{2024_Xue_PDMPure} & \cite{2025_Shidoto_AdverseCleaner} & & & & +T600 & +LM & \\
    \midrule
    Clean & 0.44 & 0.43 & 0.24 & 0.44 & 0.41 & 0.44 & 0.45 & 0.45 & 0.43 & \textbf{0.47} \\
    \midrule
    ADM+\cite{2023_Liang_AdvDM} & 0.10 & 0.14 & 0.24 & 0.13 & 0.28 & 0.21 & 0.36 & 0.36 & \textbf{0.43} & \textbf{0.43} \\
    ADM- & 0.29 & 0.27 & 0.21 & 0.31 & 0.34 & 0.25 & 0.41 & 0.42 & 0.43 & \textbf{0.44} \\
    SDS+\cite{2024_Xue_SDS} & 0.11 & 0.16 & 0.22 & 0.12 & 0.23 & 0.18 & 0.33 & 0.36 & 0.42 & \textbf{0.44} \\
    SDS-\cite{2024_Xue_SDS} & 0.25 & 0.25 & 0.20 & 0.30 & 0.30 & 0.21 & 0.39 & 0.40 & 0.41 & \textbf{0.45} \\
    EA\cite{2023_Salman_Photoguard} & 0.08 & 0.09 & 0.19 & 0.16 & 0.28 & 0.13 & 0.30 & 0.40 & 0.43 & \textbf{0.44} \\
    DA\cite{2023_Salman_Photoguard} & 0.27 & 0.26 & 0.23 & 0.28 & 0.33 & 0.25 & 0.38 & 0.40 & 0.42 & \textbf{0.47} \\
    \midrule
    Psn Avg & 0.18 & 0.19 & 0.22 & 0.21 & 0.29 & 0.21 & 0.36 & 0.39 & 0.42 & \textbf{0.45} \\
    \bottomrule
  \end{tabular}
  \label{tab:NC10_Baseline_DINOv2}
\end{table*}

\begin{table*}
  \centering
  \caption{NovelConcepts10 FID for various poison defenses.}
  \begin{tabular}{l|c|cccc|ccccc}
    \toprule
    Defense $\rightarrow$ & Nominal & Regen & PDMPure & AdvClean & JPEG & T600 & LM & JPEG & JPEG & SZT \\
    Poison $\downarrow$ & & \cite{2024_Zhao_Regeneration} & \cite{2024_Xue_PDMPure} & \cite{2025_Shidoto_AdverseCleaner} & & & & +T600 & +LM & \\
    \midrule
    Clean & 285 & 285 & 380 & 281 & 294 & 285 & 285 & 280 & 291 & \textbf{276} \\
    \midrule
    ADM+\cite{2023_Liang_AdvDM} & 453 & 416 & 378 & 443 & 349 & 389 & 309 & 311 & 291 & \textbf{280} \\
    ADM- & 339 & 353 & 390 & 336 & 324 & 361 & 299 & 294 & 293 & \textbf{279} \\
    SDS+\cite{2024_Xue_SDS} & 439 & 410 & 389 & 433 & 374 & 399 & 341 & 317 & 299 & \textbf{283} \\
    SDS-\cite{2024_Xue_SDS} & 352 & 366 & 392 & 338 & 335 & 379 & 304 & 300 & 296 & \textbf{277} \\
    EA\cite{2023_Salman_Photoguard} & 443 & 450 & 394 & 410 & 349 & 414 & 359 & 308 & 296 & \textbf{282} \\
    DA\cite{2023_Salman_Photoguard} & 352 & 350 & 373 & 354 & 323 & 367 & 309 & 305 & 299 & \textbf{269} \\
    \midrule
    Psn Avg & 396 & 391 & 386 & 385 & 342 & 385 & 320 & 306 & 296 & \textbf{278} \\
    \bottomrule
  \end{tabular}
  \label{tab:NC10_Baseline_FID}
\end{table*}

\begin{table*}
  \centering
  \caption{NovelConcepts10 CLIP Score for various poison defenses.}
  \begin{tabular}{l|c|cccc|ccccc}
    \toprule
    Defense $\rightarrow$ & Nominal & Regen & PDMPure & AdvClean & JPEG & T600 & LM & JPEG & JPEG & SZT \\
    Poison $\downarrow$ & & \cite{2024_Zhao_Regeneration} & \cite{2024_Xue_PDMPure} & \cite{2025_Shidoto_AdverseCleaner} & & & & +T600 & +LM & \\
    \midrule
    Clean & 0.51 & 0.47 & 0.49 & \textbf{0.53} & 0.48 & 0.49 & 0.52 & 0.51 & 0.50 & 0.52 \\
    \midrule
    ADM+\cite{2023_Liang_AdvDM} & 0.46 & 0.47 & 0.49 & 0.46 & 0.47 & 0.50 & 0.47 & 0.50 & 0.51 & \textbf{0.52} \\
    ADM- & 0.45 & 0.50 & 0.46 & 0.50 & 0.46 & 0.45 & \textbf{0.54} & 0.51 & 0.50 & 0.49 \\
    SDS+\cite{2024_Xue_SDS} & 0.44 & 0.49 & 0.47 & 0.48 & 0.47 & 0.48 & 0.48 & 0.48 & \textbf{0.50} & 0.49 \\
    SDS-\cite{2024_Xue_SDS} & 0.45 & 0.48 & 0.48 & 0.48 & 0.48 & 0.44 & 0.49 & 0.51 & 0.51 & \textbf{0.52} \\
    EA\cite{2023_Salman_Photoguard} & 0.45 & 0.45 & 0.47 & 0.48 & 0.48 & 0.46 & \textbf{0.50} & 0.48 & 0.47 & 0.48 \\
    DA\cite{2023_Salman_Photoguard} & 0.49 & 0.49 & 0.48 & 0.49 & 0.48 & 0.52 & 0.48 & 0.52 & \textbf{0.53} & 0.46 \\
    \midrule
    Psn Avg & 0.46 & 0.48 & 0.48 & 0.48 & 0.47 & 0.47 & 0.49 & \textbf{0.50} & \textbf{0.50} & 0.49 \\
    \bottomrule
  \end{tabular}
  \label{tab:NC10_Baseline_CLIPScore}
\end{table*}


\clearpage
\section{JPEG Preprocessing for Other Defenses}
\label{sec:appendix_JPEG_preproc}

Given the success of JPEG with SZT in Sec. \ref{subsubsec:SZT_Performance}, we further analyze JPEG preprocessing for existing defenses in Tables \ref{tab:NC10_DINO_JPEG_Preproc}, \ref{tab:NC10_FID_JPEG_Preproc}, and \ref{tab:NC10_CLIPScore_JPEG_Preproc}. In general, adding JPEG preprocessing improves performance for Regeneration and AdverseCleaner, but reduces performance for PDMPure. SZT still outperforms all baseline defenses with JPEG processing.

\section{Additional Diffusion Models}
\label{sec:appendix_Other_Models}

We further validate the effectiveness of SZT and existing defenses for TI applied to multiple LDMs. We analyze Stable Diffusion 1.5 (rectified flow version) \cite{2024_Liu_InstaFlow}, Stable Diffusion 2.1 (noise-prediction), Stable Diffusion 2.1 (velocity-prediction) \cite{2024_Lin_0SNR}, and SDXL \cite{2023_Podell_SDXL}. For convenience, we abbreviate the models as SD15, SD15rf, SD21, SD21v, and SDXL. We evaluate all models on NovelConcepts10 for all poisons. Training settings are generally similar to those described in Sec. \ref{subsubsec:Model_Training} except that we use $2500$ training steps. Due to the large size of SDXL, we instead use $500$ training steps, each with $5$ steps of gradient accumulation. We were unable to track CLIP Score for SD21, SD21v, or SDXL due to their text encoders' incompatibilities with the OpenAI CLIP image encoder.

This experiment approaches a ``black-box" poisoning scenario, as we craft all poisons using SD15 and then test defenses on different Stable Diffusion versions. This is not a major concern, as multiple prior works have demonstrated the strong transferability of poisons across Stable Diffusion models \cite{2024_Xue_SDS, 2023_Le_AntiDB}. In multiple tables for SD15rf, SD21, and SD21v results below, the ``Nominal" defense column indeed shows that all poisons are effective. Network-based defenses (Regen and PDMPure) shown here still use the same models for purification.

We evaluate SD15rf in Tables \ref{tab:NC10_SD15rfm_DINOv2}, \ref{tab:NC10_SD15rfm_FID}, and \ref{tab:NC10_SD15rfm_CLIPScore}. We suspect that the poor performance of SD15rf is due to the fact that it derives from a reflow procedure \cite{2022_Liu_RectifiedFlow}. We hypothesize that reflow, which relies on deterministic sampling for data/noise pairs to finetune models, acts as a type of model distillation and forces models to unlearn natural ODE paths between probability distributions. The rigid ODE paths enforced by distilled models may lack the flexibility to insert new concepts into the model via personalization.  Existing defenses like Regen and AdvClean barely improve the poison performance. Even so, SZT improves generative quality for TI on poisoned data to match that of clean data. 

Poison and defense performance on SD21 in Tables \ref{tab:NC10_SD21b_DINOv2} and \ref{tab:NC10_SD21b_FID} is generally similar to that of SD15, which is sensible given their similar architectures, training data, and training objectives. The trends for SD21v in Tables \ref{tab:NC10_SD21v_DINOv2} and \ref{tab:NC10_SD21v_FID} are generally similar to those of SD15 and SD21, though the effect of poisoning (i.e., the difference between ``Clean" and ``Psn Avg" values) is diminished relative to SD15 and SD21. Even so, SZT and its ablations are still effective for improving generation quality on poisoned data. In Tables \ref{tab:NC10_SDXL_DINOv2} and \ref{tab:NC10_SDXL_FID}, SDXL demonstrates the largest rift in performance trends relative to SD15, being unaffected by most poisons. Though SZT and its ablations can improve performance on poison data they are generally not necessary. We are uncertain of the exact cause of the low degree of poison transferability from SD15 to SDXL.

\section{Additional Personalization Methods}
\label{sec:appendix_Other_Personalization}

In addition to TI, we investigate additional personalization methods for Stable Diffusion 1.5, namely LoRA \cite{2021_Hu_LoRA} and CustomDiffusion \cite{2023_Kumari_CustomDiffusion}. We use NovelConcepts10 for these experiments. LoRA personalizes LDMs by finetuning low-rank adapters for U-Net weights (and optionally for the text encoder). LoRA does not finetune any text embeddings, though it does associate finetuned weights with a target prompt. CustomDiffusion is a compact version of DreamBooth \cite{2023_Ruiz_DreamBooth} that only finetunes the cross-attention weights of the U-Net as well as a new text token (as in TI). Notably, the CustomDiffusion paper also introduced a crop/rescale augmentation that applies as a mask during loss calculation, similar to our LM strategy.

In our implementation of LoRA, we use rank $4$ adapters with a target prompt that includes ``*" as a dummy token to signify LoRA usage and trained for $1000$ steps. We did not find any improvement in generative quality when using the prior preservation loss or when training text encoder weights, and thus our implementation does not include these methods. We report LoRA results in Tables \ref{tab:NC10_LoRA_DINOv2}, \ref{tab:NC10_LoRA_FID}, and \ref{tab:NC10_LoRA_CLIPScore}. We find that poisons are moderately effective against LoRA, with EA being the most severe. Existing defenses like PDMPure and AdvClean slightly improve defense against poisons but Regen generally reduces generative quality. SZT and its ablations outperform existing defenses to achieve clean-level generative quality for poisoned images. We note that as LoRA does not provide any finetuned text tokens, the CLIP Score becomes a metric for background fidelity; the CLIP text encoder will ignore the dummy token (``*") and instead focus on context words.

In our implementation of CustomDiffusion, we train for $1000$ steps and do not use the prior preservation loss. We report results both without their novel crop/rescale augmentation (Tables \ref{tab:NC10_CDnoaug_DINOv2}, \ref{tab:NC10_CDnoaug_FID}, and \ref{tab:NC10_CDnoaug_CLIPScore}) and with it (Tables \ref{tab:NC10_CDaug_DINOv2}, \ref{tab:NC10_CDaug_FID}, and \ref{tab:NC10_CDaug_CLIPScore}). Without crop/rescale, performance is generally similar to that of LoRA, with SZT outperforming existing defenses to improve poison performance. With crop/rescale, all poisons are almost ineffective and most defenses (with the exception of JPEG+T600) only reduce generative performance. This suggests that their novel crop/rescale augmentation, which is applied during loss calculation, acts similarly to our LM method, effectively blocking out adversarial signals outside the novel concept region.

\clearpage

\begin{table*}
  \centering
  \caption{NovelConcepts10 DINOv2 Similarity for existing defenses with JPEG preprocessing.}
  \begin{tabular}{l|c|cccc|ccc|c}
    \toprule
    Defense $\rightarrow$ & Nominal & Regen & PDMPure & AdvClean & JPEG & JPEG + & JPEG + & JPEG + & SZT \\
    Poison $\downarrow$ & & \cite{2024_Zhao_Regeneration} & \cite{2024_Xue_PDMPure} & \cite{2025_Shidoto_AdverseCleaner} & & Regen & PDMPure & AdvClean \\
    \midrule
    Clean & 0.44 & 0.43 & 0.24 & 0.45 & 0.41 & 0.37 & 0.19 & 0.38 & \textbf{0.47} \\
    \midrule
    ADM+\cite{2023_Liang_AdvDM} & 0.10 & 0.14 & 0.24 & 0.13 & 0.28 & 0.26 & 0.20 & 0.30 & \textbf{0.43} \\
    ADM- & 0.29 & 0.27 & 0.21 & 0.31 & 0.34 & 0.35 & 0.19 & 0.33 & \textbf{0.44} \\
    SDS+\cite{2024_Xue_SDS} & 0.11 & 0.16 & 0.22 & 0.12 & 0.23 & 0.23 & 0.19 & 0.29 & \textbf{0.44} \\
    SDS-\cite{2024_Xue_SDS} & 0.25 & 0.25 & 0.20 & 0.30 & 0.30 & 0.35 & 0.19 & 0.35 & \textbf{0.45} \\
    EA\cite{2023_Salman_Photoguard} & 0.08 & 0.09 & 0.19 & 0.16 & 0.28 & 0.29 & 0.22 & 0.32 & \textbf{0.44} \\
    DA\cite{2023_Salman_Photoguard} & 0.27 & 0.26 & 0.23 & 0.28 & 0.33 & 0.33 & 0.17 & 0.35 & \textbf{0.47} \\
    \midrule
    Psn Avg & 0.18 & 0.19 & 0.22 & 0.21 & 0.29 & 0.30 & 0.19 & 0.32 & \textbf{0.45} \\
    \bottomrule
  \end{tabular}
  \label{tab:NC10_DINO_JPEG_Preproc}
\end{table*}

\begin{table*}
  \centering
  \caption{NovelConcepts10 FID for existing defenses with JPEG preprocessing.}
  \begin{tabular}{l|c|cccc|ccc|c}
    \toprule
    Defense $\rightarrow$ & Nominal & Regen & PDMPure & AdvClean & JPEG & JPEG + & JPEG + & JPEG + & SZT \\
    Poison $\downarrow$ & & \cite{2024_Zhao_Regeneration} & \cite{2024_Xue_PDMPure} & \cite{2025_Shidoto_AdverseCleaner} & & Regen & PDMPure & AdvClean \\
    \midrule
    Clean & 285 & 285 & 380 & 281 & 294 & 303 & 389 & 303 & \textbf{276} \\
    \midrule
    ADM+\cite{2023_Liang_AdvDM} & 453 & 416 & 378 & 443 & 349 & 355 & 389 & 335 & \textbf{280} \\
    ADM- & 339 & 353 & 390 & 336 & 324 & 303 & 387 & 315 & \textbf{279} \\
    SDS+\cite{2024_Xue_SDS} & 439 & 410 & 389 & 433 & 374 & 364 & 388 & 335 & \textbf{283} \\
    SDS-\cite{2024_Xue_SDS} & 352 & 366 & 392 & 338 & 335 & 310 & 398 & 309 & \textbf{277} \\
    EA\cite{2023_Salman_Photoguard} & 443 & 450 & 394 & 410 & 349 & 335 & 370 & 321 & \textbf{282} \\
    DA\cite{2023_Salman_Photoguard} & 352 & 350 & 373 & 354 & 323 & 317 & 396 & 311 & \textbf{269} \\
    \midrule
    Psn Avg & 396 & 391 & 386 & 385 & 342 & 331 & 388 & 321 & \textbf{278} \\
    \bottomrule
  \end{tabular}
  \label{tab:NC10_FID_JPEG_Preproc}
\end{table*}

\begin{table*}
  \centering
  \caption{NovelConcepts10 CLIP Score for existing defenses with JPEG preprocessing.}
  \begin{tabular}{l|c|cccc|ccc|c}
    \toprule
    Defense $\rightarrow$ & Nominal & Regen & PDMPure & AdvClean & JPEG & JPEG + & JPEG + & JPEG + & SZT \\
    Poison $\downarrow$ & & \cite{2024_Zhao_Regeneration} & \cite{2024_Xue_PDMPure} & \cite{2025_Shidoto_AdverseCleaner} & & Regen & PDMPure & AdvClean \\
    \midrule
    Clean & 0.51 & 0.47 & 0.49 & \textbf{0.53} & 0.48 & 0.49 & 0.46 & 0.52 & 0.52 \\
    \midrule
    ADM+\cite{2023_Liang_AdvDM} & 0.46 & 0.47 & 0.49 & 0.46 & 0.47 & 0.49 & 0.51 & 0.47 & \textbf{0.52} \\
    ADM- & 0.45 & \textbf{0.50} & 0.46 & \textbf{0.50} & 0.46 & 0.48 & 0.49 & \textbf{0.50} & 0.49 \\
    SDS+\cite{2024_Xue_SDS} & 0.44 & 0.49 & 0.47 & 0.48 & 0.47 & 0.50 & \textbf{0.51} & 0.47 & 0.49 \\
    SDS-\cite{2024_Xue_SDS} & 0.45 & 0.48 & 0.48 & 0.48 & 0.48 & 0.50 & 0.49 & 0.51 & \textbf{0.52} \\
    EA\cite{2023_Salman_Photoguard} & 0.45 & 0.45 & 0.47 & 0.48 & 0.48 & \textbf{0.49} & \textbf{0.49} & 0.48 & 0.48 \\
    DA\cite{2023_Salman_Photoguard} & \textbf{0.49} & \textbf{0.49} & 0.48 & \textbf{0.49} & 0.48 & 0.48 & 0.44 & \textbf{0.49} & 0.46 \\
    \midrule
    Psn Avg & 0.46 & 0.48 & 0.48 & 0.48 & 0.47 & \textbf{0.49} & \textbf{0.49} & \textbf{0.49} & \textbf{0.49} \\
    \bottomrule
  \end{tabular}
  \label{tab:NC10_CLIPScore_JPEG_Preproc}
\end{table*}

\clearpage


\begin{table*}
  \centering
  \caption{NovelConcepts10 DINOv2 Similarity for various poison defenses with Stable Diffusion 1.5 (rectified flow).}
  \begin{tabular}{l|c|cccc|ccccc}
    \toprule
    Defense $\rightarrow$ & Nominal & Regen & PDMPure & AdvClean & JPEG & T600 & LM & JPEG & JPEG & SZT \\
    Poison $\downarrow$ & & \cite{2024_Zhao_Regeneration} & \cite{2024_Xue_PDMPure} & \cite{2025_Shidoto_AdverseCleaner} & & & & +T600 & +LM & \\
    \midrule
    Clean & 0.21 & 0.23 & 0.20 & 0.20 & 0.18 & 0.17 & \textbf{0.30} & 0.19 & 0.28 & 0.28 \\
    \midrule
    ADM+\cite{2023_Liang_AdvDM} & 0.03 & 0.04 & 0.17 & 0.02 & 0.08 & 0.18 & 0.09 & 0.16 & 0.21 & \textbf{0.24} \\
    ADM- & 0.12 & 0.13 & 0.20 & 0.13 & 0.15 & 0.12 & 0.18 & 0.14 & 0.20 & \textbf{0.23} \\
    SDS+\cite{2024_Xue_SDS} & 0.04 & 0.03 & 0.15 & 0.04 & 0.09 & 0.17 & 0.10 & 0.15 & 0.21 & \textbf{0.24} \\
    SDS-\cite{2024_Xue_SDS} & 0.13 & 0.12 & 0.18 & 0.14 & 0.12 & 0.10 & 0.17 & 0.15 & \textbf{0.22} & 0.18 \\
    EA\cite{2023_Salman_Photoguard} & 0.07 & 0.09 & 0.21 & 0.10 & 0.14 & 0.10 & 0.14 & 0.13 & \textbf{0.22} & 0.19 \\
    DA\cite{2023_Salman_Photoguard} & 0.10 & 0.08 & 0.17 & 0.10 & 0.15 & 0.11 & 0.19 & 0.16 & 0.21 & \textbf{0.23} \\
    \midrule
    Psn Avg & 0.08 & 0.08 & 0.18 & 0.09 & 0.12 & 0.13 & 0.14 & 0.15 & 0.21 & \textbf{0.22} \\
    \bottomrule
  \end{tabular}
  \label{tab:NC10_SD15rfm_DINOv2}
\end{table*}

\begin{table*}
  \centering
  \caption{NovelConcepts10 FID for various poison defenses with Stable Diffusion 1.5 (rectified flow).}
  \begin{tabular}{l|c|cccc|ccccc}
    \toprule
    Defense $\rightarrow$ & Nominal & Regen & PDMPure & AdvClean & JPEG & T600 & LM & JPEG & JPEG & SZT \\
    Poison $\downarrow$ & & \cite{2024_Zhao_Regeneration} & \cite{2024_Xue_PDMPure} & \cite{2025_Shidoto_AdverseCleaner} & & & & +T600 & +LM & \\
    \midrule
    Clean & 385 & 364 & 380 & 386 & 391 & 389 & \textbf{343} & 385 & 356 & 352 \\
    \midrule
    ADM+\cite{2023_Liang_AdvDM} & 440 & 443 & 394 & 455 & 419 & 388 & 416 & 396 & 379 & \textbf{367} \\
    ADM- & 415 & 408 & 381 & 408 & 407 & 403 & 393 & 402 & 383 & \textbf{370} \\
    SDS+\cite{2024_Xue_SDS} & 440 & 442 & 401 & 438 & 426 & 398 & 417 & 400 & 367 & \textbf{364} \\
    SDS-\cite{2024_Xue_SDS} & 407 & 416 & 397 & 409 & 409 & 410 & 395 & 398 & \textbf{371} & 399 \\
    EA\cite{2023_Salman_Photoguard} & 430 & 429 & 384 & 421 & 399 & 419 & 409 & 408 & \textbf{375} & 387 \\
    DA\cite{2023_Salman_Photoguard} & 422 & 422 & 400 & 425 & 400 & 408 & 395 & 389 & 374 & \textbf{368} \\
    \midrule
    Psn Avg & 426 & 427 & 393 & 426 & 410 & 404 & 404 & 399 & \textbf{375} & 376 \\
    \bottomrule
  \end{tabular}
  \label{tab:NC10_SD15rfm_FID}
\end{table*}

\begin{table*}
  \centering
  \caption{NovelConcepts10 CLIP Score for various poison defenses with Stable Diffusion 1.5 (rectified flow).}
  \begin{tabular}{l|c|cccc|ccccc}
    \toprule
    Defense $\rightarrow$ & Nominal & Regen & PDMPure & AdvClean & JPEG & T600 & LM & JPEG & JPEG & SZT \\
    Poison $\downarrow$ & & \cite{2024_Zhao_Regeneration} & \cite{2024_Xue_PDMPure} & \cite{2025_Shidoto_AdverseCleaner} & & & & +T600 & +LM & \\
    \midrule
    Clean & 0.44 & 0.41 & 0.46 & 0.43 & 0.42 & 0.46 & \textbf{0.47} & 0.43 & \textbf{0.47} & 0.44 \\
    \midrule
    ADM+\cite{2023_Liang_AdvDM} & 0.40 & 0.41 & 0.45 & 0.38 & 0.43 & 0.42 & 0.40 & 0.42 & 0.46 & \textbf{0.48} \\
    ADM- & 0.41 & 0.44 & \textbf{0.46} & 0.43 & 0.44 & 0.45 & \textbf{0.46} & 0.43 & 0.45 & 0.45 \\
    SDS+\cite{2024_Xue_SDS} & 0.37 & 0.37 & 0.44 & 0.39 & 0.42 & 0.45 & 0.40 & 0.44 & \textbf{0.47} & 0.46 \\
    SDS-\cite{2024_Xue_SDS} & 0.39 & 0.45 & 0.45 & 0.45 & 0.42 & 0.44 & \textbf{0.47} & 0.44 & \textbf{0.47} & 0.46 \\
    EA\cite{2023_Salman_Photoguard} & 0.40 & 0.42 & 0.45 & 0.44 & 0.42 & 0.43 & 0.43 & 0.44 & 0.45 & \textbf{0.46} \\
    DA\cite{2023_Salman_Photoguard} & 0.43 & 0.45 & 0.45 & 0.43 & 0.44 & 0.43 & 0.46 & 0.44 & 0.45 & \textbf{0.47} \\
    \midrule
    Psn Avg & 0.40 & 0.42 & 0.45 & 0.42 & 0.43 & 0.44 & 0.43 & 0.43 & \textbf{0.46} & \textbf{0.46} \\
    \bottomrule
  \end{tabular}
  \label{tab:NC10_SD15rfm_CLIPScore}
\end{table*}


\begin{table*}
  \centering
  \caption{NovelConcepts10 DINOv2 Similarity for various poison defenses with Stable Diffusion 2.1 (noise-prediction).}
  \begin{tabular}{l|c|cccc|ccccc}
    \toprule
    Defense $\rightarrow$ & Nominal & Regen & PDMPure & AdvClean & JPEG & T600 & LM & JPEG & JPEG & SZT \\
    Poison $\downarrow$ & & \cite{2024_Zhao_Regeneration} & \cite{2024_Xue_PDMPure} & \cite{2025_Shidoto_AdverseCleaner} & & & & +T600 & +LM & \\
    \midrule
    Clean & 0.41 & 0.39 & 0.31 & 0.41 & 0.38 & 0.43 & \textbf{0.44} & 0.41 & \textbf{0.44} & 0.43 \\
    \midrule
    ADM+\cite{2023_Liang_AdvDM} & 0.10 & 0.14 & 0.29 & 0.10 & 0.30 & 0.20 & 0.34 & 0.38 & 0.40 & \textbf{0.44} \\
    ADM- & 0.21 & 0.29 & 0.26 & 0.33 & 0.34 & 0.28 & 0.42 & 0.37 & \textbf{0.44} & 0.40 \\
    SDS+\cite{2024_Xue_SDS} & 0.08 & 0.13 & 0.24 & 0.10 & 0.32 & 0.18 & 0.36 & 0.32 & \textbf{0.44} & 0.43 \\
    SDS-\cite{2024_Xue_SDS} & 0.23 & 0.24 & 0.24 & 0.26 & 0.31 & 0.22 & 0.41 & 0.37 & 0.42 & \textbf{0.43} \\
    EA\cite{2023_Salman_Photoguard} & 0.09 & 0.07 & 0.26 & 0.10 & 0.26 & 0.12 & 0.33 & 0.38 & \textbf{0.43} & 0.41 \\
    DA\cite{2023_Salman_Photoguard} & 0.27 & 0.23 & 0.26 & 0.31 & 0.35 & 0.27 & 0.42 & 0.38 & \textbf{0.45} & 0.44 \\
    \midrule
    Psn Avg & 0.16 & 0.18 & 0.26 & 0.20 & 0.31 & 0.21 & 0.38 & 0.36 & \textbf{0.43} & 0.42 \\
    \bottomrule
  \end{tabular}
  \label{tab:NC10_SD21b_DINOv2}
\end{table*}

\begin{table*}
  \centering
  \caption{NovelConcepts10 FID for various poison defenses with Stable Diffusion 2.1 (noise-prediction).}
  \begin{tabular}{l|c|cccc|ccccc}
    \toprule
    Defense $\rightarrow$ & Nominal & Regen & PDMPure & AdvClean & JPEG & T600 & LM & JPEG & JPEG & SZT \\
    Poison $\downarrow$ & & \cite{2024_Zhao_Regeneration} & \cite{2024_Xue_PDMPure} & \cite{2025_Shidoto_AdverseCleaner} & & & & +T600 & +LM & \\
    \midrule
    Clean & 290 & 305 & 356 & 298 & 301 & 295 & \textbf{279} & 294 & 288 & 286 \\
    \midrule
    ADM+\cite{2023_Liang_AdvDM} & 440 & 421 & 355 & 429 & 339 & 393 & 319 & 306 & 295 & \textbf{276} \\
    ADM- & 377 & 342 & 365 & 320 & 320 & 341 & 291 & 313 & \textbf{283} & 303 \\
    SDS+\cite{2024_Xue_SDS} & 445 & 439 & 374 & 433 & 322 & 399 & 317 & 323 & 289 & \textbf{286} \\
    SDS-\cite{2024_Xue_SDS} & 362 & 368 & 383 & 356 & 318 & 371 & 299 & 313 & \textbf{290} & 295 \\
    EA\cite{2023_Salman_Photoguard} & 421 & 448 & 372 & 406 & 355 & 417 & 335 & 310 & \textbf{281} & 303 \\
    DA\cite{2023_Salman_Photoguard} & 350 & 370 & 374 & 340 & 310 & 361 & 299 & 313 & \textbf{280} & 291 \\
    \midrule
    Psn Avg & 399 & 398 & 370 & 380 & 327 & 380 & 310 & 313 & \textbf{286} & 292 \\
    \bottomrule
  \end{tabular}
  \label{tab:NC10_SD21b_FID}
\end{table*}


\begin{table*}
  \centering
  \caption{NovelConcepts10 DINOv2 Similarity for various poison defenses with Stable Diffusion 2.1 (velocity-prediction).}
  \begin{tabular}{l|c|cccc|ccccc}
    \toprule
    Defense $\rightarrow$ & Nominal & Regen & PDMPure & AdvClean & JPEG & T600 & LM & JPEG & JPEG & SZT \\
    Poison $\downarrow$ & & \cite{2024_Zhao_Regeneration} & \cite{2024_Xue_PDMPure} & \cite{2025_Shidoto_AdverseCleaner} & & & & +T600 & +LM & \\
    \midrule
    Clean & 0.28 & 0.31 & 0.26 & 0.34 & 0.30 & 0.29 & 0.39 & 0.31 & \textbf{0.41} & 0.39 \\
    \midrule
    ADM+\cite{2023_Liang_AdvDM} & 0.25 & 0.23 & 0.29 & 0.30 & 0.29 & 0.24 & 0.37 & 0.24 & \textbf{0.42} & 0.37 \\
    ADM- & 0.20 & 0.16 & 0.25 & 0.16 & 0.22 & 0.17 & 0.31 & 0.28 & 0.36 & \textbf{0.38} \\
    SDS+\cite{2024_Xue_SDS} & 0.21 & 0.21 & 0.31 & 0.27 & 0.25 & 0.23 & 0.33 & 0.28 & \textbf{0.40} & 0.38 \\
    SDS-\cite{2024_Xue_SDS} & 0.16 & 0.14 & 0.28 & 0.17 & 0.35 & 0.17 & 0.27 & 0.26 & \textbf{0.38} & \textbf{0.38} \\
    EA\cite{2023_Salman_Photoguard} & 0.09 & 0.09 & 0.28 & 0.21 & 0.26 & 0.13 & 0.27 & 0.26 & 0.38 & \textbf{0.40} \\
    DA\cite{2023_Salman_Photoguard} & 0.21 & 0.17 & 0.26 & 0.28 & 0.29 & 0.16 & 0.28 & 0.27 & \textbf{0.37} & \textbf{0.37} \\
    \midrule
    Psn Avg & 0.19 & 0.17 & 0.28 & 0.23 & 0.28 & 0.18 & 0.30 & 0.27 & \textbf{0.38} & \textbf{0.38} \\
    \bottomrule
  \end{tabular}
  \label{tab:NC10_SD21v_DINOv2}
\end{table*}

\begin{table*}
  \centering
  \caption{NovelConcepts10 FID for various poison defenses with Stable Diffusion 2.1 (velocity-prediction).}
  \begin{tabular}{l|c|cccc|ccccc}
    \toprule
    Defense $\rightarrow$ & Nominal & Regen & PDMPure & AdvClean & JPEG & T600 & LM & JPEG & JPEG & SZT \\
    Poison $\downarrow$ & & \cite{2024_Zhao_Regeneration} & \cite{2024_Xue_PDMPure} & \cite{2025_Shidoto_AdverseCleaner} & & & & +T600 & +LM & \\
    \midrule
    Clean & 336 & 328 & 343 & 321 & 326 & 334 & 308 & 330 & \textbf{295} & 314 \\
    \midrule
    ADM+\cite{2023_Liang_AdvDM} & 355 & 362 & 332 & 332 & 335 & 370 & 309 & 360 & \textbf{292} & 309 \\
    ADM- & 376 & 389 & 353 & 381 & 368 & 381 & 329 & 347 & \textbf{313} & 319 \\
    SDS+\cite{2024_Xue_SDS} & 393 & 385 & 327 & 332 & 341 & 367 & 323 & 350 & \textbf{294} & 309 \\
    SDS-\cite{2024_Xue_SDS} & 388 & 401 & 336 & 383 & 318 & 386 & 341 & 345 & \textbf{306} & 317 \\
    EA\cite{2023_Salman_Photoguard} & 431 & 436 & 336 & 370 & 348 & 420 & 371 & 352 & \textbf{300} & 316 \\
    DA\cite{2023_Salman_Photoguard} & 369 & 378 & 348 & 331 & 341 & 382 & 361 & 357 & \textbf{310} & 315 \\
    \midrule
    Psn Avg & 385 & 392 & 339 & 355 & 342 & 384 & 339 & 352 & \textbf{303} & 314 \\
    \bottomrule
  \end{tabular}
  \label{tab:NC10_SD21v_FID}
\end{table*}


\begin{table*}
  \centering
  \caption{NovelConcepts10 DINOv2 Similarity for various poison defenses with SDXL.}
  \begin{tabular}{l|c|cccc|ccccc}
    \toprule
    Defense $\rightarrow$ & Nominal & Regen & PDMPure & AdvClean & JPEG & T600 & LM & JPEG & JPEG & SZT \\
    Poison $\downarrow$ & & \cite{2024_Zhao_Regeneration} & \cite{2024_Xue_PDMPure} & \cite{2025_Shidoto_AdverseCleaner} & & & & +T600 & +LM & \\
    \midrule
    Clean & 0.35 & 0.33 & 0.25 & 0.33 & 0.22 & 0.41 & \textbf{0.42} & 0.34 & 0.39 & 0.39 \\
    \midrule
    ADM+\cite{2023_Liang_AdvDM} & 0.36 & 0.29 & 0.24 & 0.31 & 0.25 & 0.39 & \textbf{0.42} & 0.32 & 0.34 & 0.32 \\
    ADM- & 0.30 & 0.26 & 0.23 & 0.34 & 0.20 & \textbf{0.40} & 0.38 & 0.32 & 0.38 & 0.39 \\
    SDS+\cite{2024_Xue_SDS} & 0.34 & 0.37 & 0.27 & 0.32 & 0.29 & 0.39 & 0.41 & 0.39 & 0.38 & \textbf{0.42} \\
    SDS-\cite{2024_Xue_SDS} & 0.32 & 0.30 & 0.26 & 0.35 & 0.17 & 0.36 & \textbf{0.43} & 0.35 & 0.37 & 0.36 \\
    EA\cite{2023_Salman_Photoguard} & 0.30 & 0.16 & 0.22 & 0.30 & 0.20 & 0.35 & \textbf{0.42} & 0.30 & 0.31 & 0.33 \\
    DA\cite{2023_Salman_Photoguard} & 0.34 & 0.31 & 0.29 & 0.33 & 0.21 & 0.38 & \textbf{0.41} & 0.36 & 0.35 & 0.38 \\
    \midrule
    Psn Avg & 0.33 & 0.28 & 0.25 & 0.32 & 0.22 & 0.38 & \textbf{0.41} & 0.34 & 0.35 & 0.37 \\
    \bottomrule
  \end{tabular}
  \label{tab:NC10_SDXL_DINOv2}
\end{table*}

\begin{table*}
  \centering
  \caption{NovelConcepts10 FID for various poison defenses with SDXL.}
  \begin{tabular}{l|c|cccc|ccccc}
    \toprule
    Defense $\rightarrow$ & Nominal & Regen & PDMPure & AdvClean & JPEG & T600 & LM & JPEG & JPEG & SZT \\
    Poison $\downarrow$ & & \cite{2024_Zhao_Regeneration} & \cite{2024_Xue_PDMPure} & \cite{2025_Shidoto_AdverseCleaner} & & & & +T600 & +LM & \\
    \midrule
    Clean & 325 & 335 & 374 & 333 & 374 & 303 & \textbf{293} & 332 & 319 & 322 \\
    \midrule
    ADM+\cite{2023_Liang_AdvDM} & 314 & 340 & 383 & 342 & 358 & 319 & \textbf{311} & 326 & 327 & 329 \\
    ADM- & 341 & 351 & 382 & 330 & 370 & \textbf{307} & 316 & 347 & 314 & 310 \\
    SDS+\cite{2024_Xue_SDS} & 317 & 313 & 370 & 332 & 342 & 321 & 312 & 317 & 314 & \textbf{298} \\
    SDS-\cite{2024_Xue_SDS} & 336 & 322 & 371 & \textbf{321} & 386 & 325 & 297 & 323 & 322 & 332 \\
    EA\cite{2023_Salman_Photoguard} & 337 & 421 & 387 & 334 & 376 & 328 & \textbf{303} & 337 & 344 & 321 \\
    DA\cite{2023_Salman_Photoguard} & 325 & 334 & 356 & 331 & 378 & 325 & 319 & 318 & 324 & \textbf{313} \\
    \midrule
    Psn Avg & 328 & 347 & 375 & 332 & 368 & 321 & \textbf{310} & 328 & 324 & 317 \\
    \bottomrule
  \end{tabular}
  \label{tab:NC10_SDXL_FID}
\end{table*}

\clearpage


\begin{table*}
  \centering
  \caption{NovelConcepts10 DINOv2 Similarity for various poison defenses with LoRA.}
  \begin{tabular}{l|c|cccc|ccccc}
    \toprule
    Defense $\rightarrow$ & Nominal & Regen & PDMPure & AdvClean & JPEG & T600 & LM & JPEG & JPEG & SZT \\
    Poison $\downarrow$ & & \cite{2024_Zhao_Regeneration} & \cite{2024_Xue_PDMPure} & \cite{2025_Shidoto_AdverseCleaner} & & & & +T600 & +LM & \\
    \midrule
    Clean & 0.35 & 0.35 & 0.25 & 0.34 & 0.35 & \textbf{0.47} & 0.39 & 0.42 & 0.38 & 0.44 \\
    \midrule
    ADM+\cite{2023_Liang_AdvDM} & 0.26 & 0.26 & 0.28 & 0.28 & \textbf{0.35} & 0.28 & 0.31 & \textbf{0.35} & 0.31 & \textbf{0.35} \\
    ADM- & 0.33 & 0.29 & 0.29 & 0.35 & 0.32 & 0.28 & 0.32 & 0.40 & 0.36 & \textbf{0.42} \\
    SDS+\cite{2024_Xue_SDS} & 0.32 & 0.26 & 0.31 & 0.29 & 0.30 & 0.28 & 0.28 & 0.31 & 0.32 & \textbf{0.34} \\
    SDS-\cite{2024_Xue_SDS} & 0.31 & 0.27 & 0.28 & 0.35 & 0.32 & 0.26 & 0.31 & \textbf{0.41} & 0.35 & 0.38 \\
    EA\cite{2023_Salman_Photoguard} & 0.11 & 0.13 & 0.30 & 0.18 & 0.29 & 0.09 & 0.13 & 0.35 & 0.31 & \textbf{0.36} \\
    DA\cite{2023_Salman_Photoguard} & 0.29 & 0.27 & 0.27 & 0.32 & 0.32 & 0.24 & 0.28 & \textbf{0.40} & 0.34 & 0.38 \\
    \midrule
    Psn Avg & 0.27 & 0.25 & 0.29 & 0.30 & 0.32 & 0.24 & 0.27 & \textbf{0.37} & 0.33 & \textbf{0.37} \\
    \bottomrule
  \end{tabular}
  \label{tab:NC10_LoRA_DINOv2}
\end{table*}

\begin{table*}
  \centering
  \caption{NovelConcepts10 FID for various poison defenses with LoRA.}
  \begin{tabular}{l|c|cccc|ccccc}
    \toprule
    Defense $\rightarrow$ & Nominal & Regen & PDMPure & AdvClean & JPEG & T600 & LM & JPEG & JPEG & SZT \\
    Poison $\downarrow$ & & \cite{2024_Zhao_Regeneration} & \cite{2024_Xue_PDMPure} & \cite{2025_Shidoto_AdverseCleaner} & & & & +T600 & +LM & \\
    \midrule
    Clean & 322 & 311 & 365 & 316 & 315 & \textbf{261} & 301 & 284 & 303 & 280 \\
    \midrule
    ADM+\cite{2023_Liang_AdvDM} & 364 & 363 & 357 & 367 & 312 & 383 & 360 & \textbf{310} & 337 & 325 \\
    ADM- & 328 & 347 & 360 & 318 & 319 & 345 & 350 & \textbf{284} & 316 & 287 \\
    SDS+\cite{2024_Xue_SDS} & 335 & 355 & 348 & 346 & 339 & 375 & 363 & 340 & 330 & \textbf{323} \\
    SDS-\cite{2024_Xue_SDS} & 338 & 350 & 357 & 318 & 320 & 355 & 345 & \textbf{282} & 317 & 298 \\
    EA\cite{2023_Salman_Photoguard} & 435 & 433 & 347 & 414 & 351 & 432 & 425 & 328 & 339 & \textbf{318} \\
    DA\cite{2023_Salman_Photoguard} & 348 & 356 & 365 & 342 & 323 & 388 & 352 & \textbf{292} & 323 & 311 \\
    \midrule
    Psn Avg & 358 & 367 & 356 & 351 & 327 & 380 & 366 & \textbf{306} & 327 & 310 \\
    \bottomrule
  \end{tabular}
  \label{tab:NC10_LoRA_FID}
\end{table*}

\begin{table*}
  \centering
  \caption{NovelConcepts10 CLIP Score for various poison defenses with LoRA.}
  \begin{tabular}{l|c|cccc|ccccc}
    \toprule
    Defense $\rightarrow$ & Nominal & Regen & PDMPure & AdvClean & JPEG & T600 & LM & JPEG & JPEG & SZT \\
    Poison $\downarrow$ & & \cite{2024_Zhao_Regeneration} & \cite{2024_Xue_PDMPure} & \cite{2025_Shidoto_AdverseCleaner} & & & & +T600 & +LM & \\
    \midrule
    Clean & 0.48 & 0.48 & \textbf{0.51} & 0.48 & 0.48 & 0.41 & 0.44 & 0.44 & 0.45 & 0.42 \\
    \midrule
    ADM+\cite{2023_Liang_AdvDM} & 0.49 & 0.49 & \textbf{0.50} & \textbf{0.50} & 0.48 & 0.46 & 0.45 & 0.45 & 0.48 & 0.45 \\
    ADM- & 0.47 & 0.48 & 0.48 & 0.47 & \textbf{0.49} & 0.42 & 0.46 & 0.43 & 0.46 & 0.43 \\
    SDS+\cite{2024_Xue_SDS} & 0.48 & 0.49 & 0.47 & \textbf{0.50} & 0.49 & 0.45 & 0.47 & 0.46 & 0.48 & 0.45 \\
    SDS-\cite{2024_Xue_SDS} & 0.48 & 0.46 & 0.47 & 0.48 & \textbf{0.49} & 0.44 & 0.45 & 0.44 & 0.46 & 0.44 \\
    EA\cite{2023_Salman_Photoguard} & 0.42 & 0.42 & 0.46 & 0.43 & \textbf{0.48} & 0.41 & 0.43 & 0.44 & 0.47 & 0.45 \\
    DA\cite{2023_Salman_Photoguard} & 0.47 & 0.48 & 0.48 & 0.48 & \textbf{0.49} & 0.44 & 0.47 & 0.44 & 0.47 & 0.45 \\
    \midrule
    Psn Avg & 0.47 & 0.47 & \textbf{0.48} & \textbf{0.48} & \textbf{0.48} & 0.44 & 0.45 & 0.44 & 0.47 & 0.44 \\
    \bottomrule
  \end{tabular}
  \label{tab:NC10_LoRA_CLIPScore}
\end{table*}


\begin{table*}
  \centering
  \caption{NovelConcepts10 DINOv2 Similarity for various poison defenses with CustomDiffusion (no crop/rescale augmentation).}
  \begin{tabular}{l|c|cccc|ccccc}
    \toprule
    Defense $\rightarrow$ & Nominal & Regen & PDMPure & AdvClean & JPEG & T600 & LM & JPEG & JPEG & SZT \\
    Poison $\downarrow$ & & \cite{2024_Zhao_Regeneration} & \cite{2024_Xue_PDMPure} & \cite{2025_Shidoto_AdverseCleaner} & & & & +T600 & +LM & \\
    \midrule
    Clean & 0.30 & 0.31 & 0.27 & 0.30 & 0.30 & \textbf{0.37} & 0.30 & 0.34 & 0.32 & \textbf{0.37} \\
    \midrule
    ADM+\cite{2023_Liang_AdvDM} & 0.26 & 0.28 & 0.29 & 0.28 & 0.30 & 0.27 & 0.31 & 0.33 & 0.34 & \textbf{0.35} \\
    ADM- & 0.29 & 0.25 & 0.27 & 0.30 & 0.31 & 0.20 & 0.29 & 0.32 & 0.32 & \textbf{0.33} \\
    SDS+\cite{2024_Xue_SDS} & 0.25 & 0.27 & 0.28 & 0.26 & 0.29 & 0.26 & 0.29 & 0.33 & 0.33 & \textbf{0.35} \\
    SDS-\cite{2024_Xue_SDS} & 0.27 & 0.25 & 0.28 & 0.30 & 0.32 & 0.19 & 0.29 & 0.32 & 0.34 & \textbf{0.38} \\
    EA\cite{2023_Salman_Photoguard} & 0.14 & 0.15 & 0.24 & 0.19 & 0.29 & 0.13 & 0.19 & 0.30 & \textbf{0.33} & \textbf{0.33} \\
    DA\cite{2023_Salman_Photoguard} & 0.28 & 0.25 & 0.27 & 0.30 & 0.31 & 0.23 & 0.31 & 0.29 & \textbf{0.34} & \textbf{0.34} \\
    \midrule
    Psn Avg & 0.25 & 0.24 & 0.27 & 0.27 & 0.30 & 0.21 & 0.28 & 0.31 & 0.33 & \textbf{0.35} \\
    \bottomrule
  \end{tabular}
  \label{tab:NC10_CDnoaug_DINOv2}
\end{table*}

\begin{table*}
  \centering
  \caption{NovelConcepts10 FID for various poison defenses with CustomDiffusion (no crop/rescale augmentation).}
  \begin{tabular}{l|c|cccc|ccccc}
    \toprule
    Defense $\rightarrow$ & Nominal & Regen & PDMPure & AdvClean & JPEG & T600 & LM & JPEG & JPEG & SZT \\
    Poison $\downarrow$ & & \cite{2024_Zhao_Regeneration} & \cite{2024_Xue_PDMPure} & \cite{2025_Shidoto_AdverseCleaner} & & & & +T600 & +LM & \\
    \midrule
    Clean & 336 & 333 & 346 & 331 & 327 & \textbf{305} & 335 & 316 & 324 & 307 \\
    \midrule
    ADM+\cite{2023_Liang_AdvDM} & 347 & 346 & 342 & 339 & 331 & 347 & 330 & 322 & 311 & \textbf{306} \\
    ADM- & 335 & 352 & 349 & 338 & 334 & 374 & 338 & 332 & 324 & \textbf{320} \\
    SDS+\cite{2024_Xue_SDS} & 357 & 345 & 341 & 354 & 334 & 350 & 341 & 319 & 325 & \textbf{309} \\
    SDS-\cite{2024_Xue_SDS} & 344 & 348 & 344 & 333 & 324 & 383 & 343 & 330 & 314 & \textbf{302} \\
    EA\cite{2023_Salman_Photoguard} & 404 & 393 & 360 & 382 & 339 & 408 & 389 & 336 & \textbf{318} & 321 \\
    DA\cite{2023_Salman_Photoguard} & 342 & 351 & 347 & 332 & 333 & 368 & 331 & 339 & 315 & \textbf{310} \\
    \midrule
    Psn Avg & 355 & 356 & 347 & 346 & 333 & 372 & 345 & 330 & 318 & \textbf{312} \\
    \bottomrule
  \end{tabular}
  \label{tab:NC10_CDnoaug_FID}
\end{table*}

\begin{table*}
  \centering
  \caption{NovelConcepts10 CLIP Score for various poison defenses with CustomDiffusion (no crop/rescale augmentation).}
  \begin{tabular}{l|c|cccc|ccccc}
    \toprule
    Defense $\rightarrow$ & Nominal & Regen & PDMPure & AdvClean & JPEG & T600 & LM & JPEG & JPEG & SZT \\
    Poison $\downarrow$ & & \cite{2024_Zhao_Regeneration} & \cite{2024_Xue_PDMPure} & \cite{2025_Shidoto_AdverseCleaner} & & & & +T600 & +LM & \\
    \midrule
    Clean & 0.37 & 0.38 & 0.38 & 0.38 & 0.38 & 0.39 & 0.38 & 0.39 & \textbf{0.40} & 0.39 \\
    \midrule
    ADM+\cite{2023_Liang_AdvDM} & \textbf{0.40} & 0.39 & \textbf{0.40} & 0.39 & 0.39 & 0.39 & \textbf{0.40} & \textbf{0.40} & \textbf{0.40} & 0.39 \\
    ADM- & 0.38 & 0.39 & 0.39 & 0.38 & 0.38 & \textbf{0.41} & 0.39 & 0.38 & 0.40 & 0.39 \\
    SDS+\cite{2024_Xue_SDS} & \textbf{0.40} & 0.38 & 0.38 & \textbf{0.40} & 0.39 & \textbf{0.40} & 0.39 & 0.39 & 0.39 & \textbf{0.40} \\
    SDS-\cite{2024_Xue_SDS} & 0.37 & 0.39 & 0.38 & 0.38 & 0.39 & \textbf{0.41} & 0.40 & 0.39 & 0.39 & 0.40 \\
    EA\cite{2023_Salman_Photoguard} & 0.38 & 0.38 & 0.38 & 0.37 & 0.38 & 0.39 & \textbf{0.40} & 0.39 & 0.39 & 0.39 \\
    DA\cite{2023_Salman_Photoguard} & \textbf{0.39} & 0.38 & 0.38 & 0.38 & 0.38 & \textbf{0.39} & \textbf{0.39} & \textbf{0.39} & 0.38 & \textbf{0.39} \\
    \midrule
    Psn Avg & 0.39 & 0.39 & 0.38 & 0.38 & 0.39 & \textbf{0.40} & 0.39 & 0.39 & 0.39 & 0.39 \\
    \bottomrule
  \end{tabular}
  \label{tab:NC10_CDnoaug_CLIPScore}
\end{table*}

\begin{table*}
  \centering
  \caption{NovelConcepts10 DINOv2 Similarity for various poison defenses with CustomDiffusion (crop/rescale augmentation).}
  \begin{tabular}{l|c|cccc|ccccc}
    \toprule
    Defense $\rightarrow$ & Nominal & Regen & PDMPure & AdvClean & JPEG & T600 & LM & JPEG & JPEG & SZT \\
    Poison $\downarrow$ & & \cite{2024_Zhao_Regeneration} & \cite{2024_Xue_PDMPure} & \cite{2025_Shidoto_AdverseCleaner} & & & & +T600 & +LM & \\
    \midrule
    Clean & 0.35 & 0.35 & 0.33 & \textbf{0.36} & 0.33 & \textbf{0.36} & 0.34 & \textbf{0.36} & 0.34 & 0.32 \\
    \midrule
    ADM+\cite{2023_Liang_AdvDM} & 0.32 & 0.30 & 0.33 & \textbf{0.34} & 0.31 & \textbf{0.34} & 0.33 & 0.33 & 0.33 & 0.30 \\
    ADM- & 0.33 & 0.25 & 0.33 & 0.33 & 0.33 & 0.33 & 0.31 & \textbf{0.34} & \textbf{0.34} & 0.30 \\
    SDS+\cite{2024_Xue_SDS} & 0.33 & 0.31 & \textbf{0.34} & 0.31 & 0.32 & 0.32 & 0.32 & \textbf{0.34} & \textbf{0.34} & 0.30 \\
    SDS-\cite{2024_Xue_SDS} & 0.32 & 0.26 & 0.32 & \textbf{0.34} & 0.33 & 0.31 & 0.31 & \textbf{0.34} & 0.32 & 0.30 \\
    EA\cite{2023_Salman_Photoguard} & 0.31 & 0.21 & 0.29 & 0.31 & 0.31 & 0.29 & 0.32 & 0.32 & \textbf{0.33} & 0.30 \\
    DA\cite{2023_Salman_Photoguard} & \textbf{0.36} & 0.33 & 0.32 & 0.32 & 0.32 & 0.34 & 0.33 & 0.35 & 0.33 & 0.30 \\
    \midrule
    Psn Avg & 0.33 & 0.28 & 0.32 & 0.32 & 0.32 & 0.32 & 0.32 & \textbf{0.34} & 0.33 & 0.30 \\
    \bottomrule
  \end{tabular}
  \label{tab:NC10_CDaug_DINOv2}
\end{table*}

\begin{table*}
  \centering
  \caption{NovelConcepts10 FID for various poison defenses with CustomDiffusion (crop/rescale augmentation).}
  \begin{tabular}{l|c|cccc|ccccc}
    \toprule
    Defense $\rightarrow$ & Nominal & Regen & PDMPure & AdvClean & JPEG & T600 & LM & JPEG & JPEG & SZT \\
    Poison $\downarrow$ & & \cite{2024_Zhao_Regeneration} & \cite{2024_Xue_PDMPure} & \cite{2025_Shidoto_AdverseCleaner} & & & & +T600 & +LM & \\
    \midrule
    Clean & 316 & \textbf{313} & 321 & 315 & 323 & 314 & 320 & \textbf{313} & 320 & 325 \\
    \midrule
    ADM+\cite{2023_Liang_AdvDM} & 329 & 330 & 325 & \textbf{319} & 326 & 321 & 323 & 324 & 322 & 333 \\
    ADM- & 324 & 353 & 320 & 327 & \textbf{320} & 327 & 327 & 321 & 321 & 334 \\
    SDS+\cite{2024_Xue_SDS} & 323 & 331 & 319 & 329 & 325 & 328 & 326 & 320 & \textbf{316} & 328 \\
    SDS-\cite{2024_Xue_SDS} & 324 & 357 & 326 & 320 & \textbf{317} & 338 & 329 & 324 & 327 & 337 \\
    EA\cite{2023_Salman_Photoguard} & 332 & 375 & 336 & 335 & 331 & 349 & 324 & 331 & \textbf{323} & 337 \\
    DA\cite{2023_Salman_Photoguard} & \textbf{311} & 326 & 329 & 327 & 327 & 320 & 325 & 321 & 324 & 332 \\
    \midrule
    Psn Avg & 324 & 345 & 326 & 326 & 325 & 330 & 326 & 324 & \textbf{322} & 333 \\
    \bottomrule
  \end{tabular}
  \label{tab:NC10_CDaug_FID}
\end{table*}

\begin{table*}
  \centering
  \caption{NovelConcepts10 CLIP Score for various poison defenses with CustomDiffusion (crop/rescale augmentation).}
  \begin{tabular}{l|c|cccc|ccccc}
    \toprule
    Defense $\rightarrow$ & Nominal & Regen & PDMPure & AdvClean & JPEG & T600 & LM & JPEG & JPEG & SZT \\
    Poison $\downarrow$ & & \cite{2024_Zhao_Regeneration} & \cite{2024_Xue_PDMPure} & \cite{2025_Shidoto_AdverseCleaner} & & & & +T600 & +LM & \\
    \midrule
    Clean & 0.41 & 0.40 & \textbf{0.42} & 0.41 & 0.40 & 0.40 & \textbf{0.42} & 0.41 & \textbf{0.42} & 0.40 \\
    \midrule
    ADM+\cite{2023_Liang_AdvDM} & 0.41 & 0.41 & 0.41 & 0.40 & 0.41 & 0.41 & 0.42 & 0.41 & \textbf{0.43} & 0.41 \\
    ADM- & 0.42 & 0.42 & 0.41 & 0.40 & 0.41 & 0.40 & 0.42 & 0.39 & \textbf{0.43} & 0.42 \\
    SDS+\cite{2024_Xue_SDS} & 0.42 & \textbf{0.43} & 0.42 & 0.40 & \textbf{0.43} & 0.41 & 0.42 & 0.40 & 0.42 & 0.42 \\
    SDS-\cite{2024_Xue_SDS} & \textbf{0.42} & 0.40 & 0.41 & 0.41 & 0.41 & \textbf{0.42} & 0.41 & 0.41 & \textbf{0.42} & 0.41 \\
    EA\cite{2023_Salman_Photoguard} & 0.41 & 0.40 & \textbf{0.42} & 0.41 & 0.40 & 0.41 & 0.40 & 0.40 & \textbf{0.42} & 0.41 \\
    DA\cite{2023_Salman_Photoguard} & 0.41 & \textbf{0.42} & \textbf{0.42} & 0.40 & \textbf{0.42} & 0.41 & \textbf{0.42} & 0.40 & \textbf{0.42} & 0.40 \\
    \midrule
    Psn Avg & 0.41 & 0.41 & \textbf{0.42} & 0.40 & 0.41 & 0.41 & \textbf{0.42} & 0.40 & \textbf{0.42} & 0.41 \\
    \bottomrule
  \end{tabular}
  \label{tab:NC10_CDaug_CLIPScore}
\end{table*}

\clearpage
\section{Ethical Statement}

As our research primarily concerns the subfield of data poisoning, we are keenly aware of our work's ethical proximity to copyright theft and artistic style copying. Practical applications of SZT (and related methods) may realize as improved attacks against attempts by copyright holders and artists to protect their works. Even so, we believe that our research is necessary. The individual components of SZT are not complex, simply relying on JPEG compression, biased timestep sampling, and loss masking. Rather, we believe that the effectiveness of SZT, despite its simplicity, exposes the vulnerabilities of existing poisons and demands further research on robust poisons.

\section{Contributions}

Name order generally denotes share of contribution.

\noindent Dataset Collection: Styborski, Kapur, Lu

\noindent Poison Curation: Styborski, Kapur

\noindent Defense Baselines: Lu, Kapur, Styborski

\noindent Semantic Sensitivity Map Studies: Lyu

\noindent Timestep Learning Bias Studies: Styborski

\noindent Spatial Learning Bias Studies: Lyu, Styborski

\noindent JPEG Compression Studies: Styborski

\noindent Hyperparameter Ablation: Styborski

\noindent Model Ablation: Lyu, Styborski, Lu

\noindent Additional Personalization Methods: Styborski

\noindent Writing: Styborski, Lyu, Kong

\noindent Editing: Kong, Styborski, Lyu

\noindent Advising and Guidance: Kong

\end{document}